\documentclass{aa}  
\def\sqr#1#2{{\vcenter{\vbox{\hrule height.#2pt
    \hbox{\vrule width.#2pt height#1pt \kern#1pt \vrule width.#2pt}
    \hrule height.#2pt}}}}

\usepackage{pifont}
\usepackage{amssymb}
\usepackage{wasysym}
\usepackage[dvips]{epsfig}
\usepackage{graphicx}
\usepackage{ctable}
\usepackage{deleq}
\usepackage{natbib}
\usepackage{multirow}

\begin{document}

   \title{Combined effects of tidal and rotational distortions on the
   equilibrium  configuration of low-mass, pre-main sequence stars%
   \thanks{The complete version of Table\,\ref{tab100sm00} is only 
   available in electronic form at the CDS via anonymous ftp to 
   cdsarc.u-strasbg.fr (130.79.128.5) or via 
   http://cdsweb.u-strasbg.fr/cgi-bin/qcat?J/A+A/.}}


   \author{N.R. Landin\inst{1},  
      L.T.S. Mendes\inst{1,2} \and L.P.R. Vaz\inst{1} 
          }

   \offprints{N.R.Landin}

   \institute{Depto.\ de F\'{\i}sica,
              Universidade Federal de Minas Gerais, C.P.702, 31270-901 --
              Belo Horizonte, MG, Brazil; \\
              \email{nlandin@fisica.ufmg.br, lpv@fisica.ufmg.br}
              \and
              Depto.\ de Engenharia Eletr\^onica,
              Universidade Federal de Minas Gerais, C.P.702, 31270-901 --
              Belo Horizonte, MG, Brazil; \\
              \email{luizt@cpdee.ufmg.br}  
             }

   \date{Received \dots; accepted \dots}

 
\abstract
{
In close binary systems, the axial rotation and the mutual tidal forces
of the component stars deform each other and destroy their spherical
symmetry by means of the respective disturbing potentials. 
}
{
We present new models for low-mass, pre-main sequence stars that include 
the combined distortion effects of tidal and rotational forces
on the equilibrium configuration of stars. Using our theoretical results, we aim
at investigating the effects of interaction between tides and rotation 
on the stellar structure and evolution. 
}
{
The Kippenhahn \& Thomas (1970)
approximation, along with the Clairaut-Legendre expansion for the
gravitational potential of a self-gravitating body, is used
to take the effects of tidal and rotational distortions on the
stellar configuration into account.
}
{
We obtained values of internal structure constants for low-mass, pre-main sequence
stars from stellar evolutionary models that consider the
combined effects of rotation and tidal forces due to a companion star.
We also derived a new expression for the rotational inertia of a tidally and
rotationally distorted star. Our values corresponding to standard models (with no distortions)
are compatible with those available in literature.
Our distorted models were successfully used to analyze the
eclipsing binary system EK Cep, reproducing the stellar radii, 
effective temperature ratio, lithium depletion, rotational velocities, 
and the apsidal motion rate in the age interval of 15.5-16.7\,Myr. 
}
{
In the low-mass range, the assumption that harmonics 
greater than $j$=2 can be neglected seems not to be fully justified,
although it is widely used when analyzing the
apsidal motion of binary systems.
The non-standard evolutionary tracks are cooler than
the standard ones, mainly for low-mass stars. Distorted models predict
more mass-concentrated stars at the zero-age main-sequence than standard 
models.
}


\authorrunning {Landin et al.}
\titlerunning {Tidal and/or rotational distortions}

\maketitle


\section{Introduction}

A binary system consists of two stars that rotate around their own
axes and, at the same time, revolve around the center of mass of the system.
Sometimes, the intrinsic rotation axis and the orbital one are aligned 
since the beginning of the formation process. Usually, the orbit
begins with a considerable eccentricity and the component stars are not 
synchronized with the orbital angular velocity. However, due to the inertial forces that take
place, the system tends to align its axes, to synchronize the rotational 
angular velocity
of the components with the orbital angular velocity at the periastron passage
and, finally, to circularize the orbit.
Extensive spectroscopic evidence reveals that the 
components of close binary (in general, non-eccentric) systems do rotate with an angular velocity, $\Omega$,
which is generally equal to the Keplerian angular velocity, $\omega_K$, of the 
orbital motion around a common center of mass, so that
\begin{equation}
\label{kepler}
\Omega \cong \omega_K = \sqrt{G{M_1+M_2 \over R^3}}.
\end{equation}

However, occasionally $\Omega$ is much larger than $\omega_K$ -- the sense of
rotation being direct in every known case. In systems exhibiting circular 
orbits, synchronism between rotation and revolution may usually (though not 
always) be expected to exist, while components describing eccentric 
orbits, though in general rotating faster than their mean orbital
angular velocity, can still be synchronized with the orbital motion at the
periastron.

When both components are on the main sequence, the most massive one
is also larger in radius and hotter, and is called primary, both in
photometric (the star eclipsed in the deepest minimum of the light
curve) and in spectroscopic studies. In other evolutionary stages,
however, it is a common situation that the most massive component is
not the larger or the one of higher effective temperature. It continues
being designed as ``primary'' in spectroscopic studies, but referred to
as ``secondary'' in light curve analysis, due to the lower effective
temperature.

One important aspect of the evolution of close binaries is the
dynamical evolution due to tidal interaction, which is reflected in 
the rotation of the stars and in the eccentricity of their orbits.
Tidal deformation due to the companion would be symmetric about the line 
joining their centers, if there were no dissipation of kinetic energy 
into heat. It is this dissipation that induces a phase shift in the 
tidal bulge, and the tilted mass distribution, then, exerts a torque on 
the star, leading to an exchange of angular momentum between its spin 
and the orbital motion. Theory distinguishes two components in the tide,
namely, equilibrium tide and dynamical tide \citep{zahn89}:
\vspace{-.5\baselineskip}
\begin{itemize}
  \advance\itemsep by -1.5\itemsep
  \item Equilibrium tide is the hydrostatic adjustment of the structure of 
        the star to the perturbing force exerted by the companion. The
        dissipation mechanism acting on this tide is the interaction
        between the convective motions and the tidal flow \citep{zahn66}.
  \item Dynamical tide is the dynamical response to the tidal force exerted
        by the companion; it takes into account the elastic properties of
        the star, and the possibilities of resonances with its free modes
        of oscillation. The dissipation mechanism acting on this tide is
        the departure from adiabaticity of the forced oscillation, due to
       the radiative damping \citep{zahn75}.
\end{itemize}
To describe the tidal process in massive main-sequence
stars, which have convective cores and radiative envelopes, it is 
necessary to use a theory that accounts for dynamical effects which 
arise due to tidal forces \citep{zahn77,savonije83}.
\citet{witte99a,witte99b,witte01} published extensive studies of
the role played by close resonances with eigenmodes
of these early type stars during the orbital decay 
\citep[see also][]{willems03I}.
On the other hand, for low-mass main-sequence stars
and giants, which have extended 
convective envelopes, retardation of the equilibrium
tide due to the viscosity of turbulent eddies in the envelope is usually
assumed to be the cause of the tidal torque.
\citet{witte02a} calculated the tidal interaction of a uniformly rotating
1\,M$_{\odot}$ star with a orbiting companion at various phases of
evolution from the zero-age main-sequence (ZAMS) to core hydrogen exhaustion. Their results
indicate that effects related to stellar rotation can considerably enhance 
the speed of tidal evolution in low-mass binary systems. In another paper,
\citet{witte02b} showed that energy dissipation through 
resonant dynamic tides may dominate convective damping of equilibrium 
tides in solar-type stars.

In standard models, the stars are assumed to be spherically
symmetric. However, the spherical symmetry will be destroyed if a 
disturbing potential exists, as it happens for rotating stars either
isolated or in binary (multiple) systems, or due to
tidal forces (gravitational influence of a companion),
present in binary systems.

In the case of a rotating star in a binary system, both rotational
and tidal forces distort its shape from the spherical symmetry.
The analytic determination of these combined effects is complex and
approximate methods have been used in the literature,
with one of the distorting forces (generally rotation) being
analyzed in approximate ways \citep{mohan90}. \citet{chandra33}
developed the theory of distorted polytropes and \citet{kopal72,kopal74}
developed the concept of Roche equipotentials and coordinates to study
the combined effects of tidal forces and rotation on stars.
\citet{kippen70} (KT70) devised a method for introducing the effects of rotation
in existing one-dimensional evolutionary codes that became widely
adopted in the literature \citep[see e.g.][]{endal76, pinson90, fliegner95,
chaboyer95, meynet97, mendes, claret99}. In the case of massive stars, the effects of
rotation on their structure and evolution were studied by various authors
(e.g. \citealt{meynet97, meynet00}; \citealt{maeder98}; \citealt{heger00};
\citealt{maeder00,maeder01}), some of which present also grids of stellar
models for rotating, massive stars.

In the low-mass range, \citet{pinson90}, \citet{claret96} and
\citet{mendes} presented exploratory studies on the influence of rotation
on Li depletion in pre-main sequence (MS) evolution, and \citet{landin06} provided grids
of non-gray rotating pre-MS models from 0.085\,M$_{\odot}$ to 3.8\,M$_{\odot}$.

\citet{mohan90} presented a method for calculating the equilibrium
structure of a rotationally and tidally distorted primary component in a
synchronously rotating binary system. Their method assumes a Roche model
potential and uses the KT70 method for computing the corrections due to
rotation and tides. Their results for stellar models
of 10, 5, and 2.5\,M$_{\odot}$, for which the mass ratio $q$ of the secondary component
to the primary component was set to 0.1, indicate that rotational effects
are more important than those produced by tidal distortions.

In this paper, we present a new version of the {\tt ATON} code 
\citep{ventura98} that treats the combined effects of tidal and rotational
distortions on a star. This allows us, for example, to obtain the apsidal
motion constants of the primary component of a binary system that rotates and
suffers the tidal effects of the secondary component.
Instead, however, of using a Roche-type potential, which considers both binary
components as point masses, we adopt the more precise
technique based on the Clairaut-Legendre expansion for the potential energy 
of a self-gravitating body \citep{kopal59}. The paper is organized as follows:
in Sect.\,(\ref{apsmotion}) we give a brief account of theoretical apsidal
motion calculations in the literature. Sect.\,(\ref{kipthoform}) presents the
KT70 method, proposed for determining the equilibrium structures of rotationally and
tidally distorted stellar models, in which the non-spherical stellar equations can
be easily obtained from the spherical ones. The technique for introducing the
combined tidal and rotational effects is described in Sect.\,(\ref{distortions}),
in which we present, also, new calculations of internal  structure constants extended
to the pre-MS phase.
The results are presented in the Sect.\,(\ref{iscresult}). Discussion and 
comparisons with observed apsidal motion rates are given in
Sect.\,(\ref{iscdisc}).

\section{Apsidal motion and internal structure constants}\label{apsmotion}

The internal structure constants
$k_2$, $k_3$ and $k_4$, also known as apsidal motion constants, are
important in stellar astrophysics. They are mass concentration parameters
that depend on the mass distribution throughout the
star. There is, however, a direct relation between the gravitational field
of a non-spherical body and the internal density concentration in that body
\citep{sahade78}.

From a theoretical point of view, the values of $k_j$ ($j$=2,3,4) depend
on the model used. For the Roche model, in which the whole stellar mass
is concentrated at its center, the $k_j$ values are all
equal to zero, while for a homogeneous model $k_2$=3/4, $k_3$=3/8 and $k_4$=1/4. 
The values of the internal structure constants are essential to compute the
theoretical apsidal motion rates in close binaries, and the comparison with
the observations constitutes an important test for evolutionary models.
The most centrally concentrated stars have the lowest values of $k_j$ and
the longest values of apsidal periods (Eq.\,\ref{k2obs}).

It can be shown that the theoretical apsidal rate $\dot \omega$,
in radians per cycle, is given in terms of the internal structure constants
by \citep[hereafter H87]{martynov73, hejlesen87}
\begin{equation}
{\dot \omega \over 2\pi}= \sum_{i=1}^2 \sum _{j=2}^4 c_{ji}k_{ji},
\label{isc234}
\end{equation}
where $i$ denotes the component star (1=primary, 2=secondary)
and $j$ the harmonic order. Generally, terms of order higher
than $j$=2 are very small so that only values of $k_{2i}$ enter in
Eq.\,(\ref{isc234}).

The internal structure constants are important in other astrophysical
aspects, since synchronization and circularization time scales in close
binaries depend on $k_2$ \citep{zahn77}. Other applications
are in the computation of rotational angular momenta
(Sect.\,\ref{rotinertia}), where gyration radii (defined in Eq.\,\ref{beta}) 
can be expressed as a linear function of the apsidal motion constants
\citep{ureche76}, and in the determination of the effect of binarity
in the geometry of the stellar surfaces due to rotation and tides
\citep{rucinski69,kopal78}.

The first analytical expression for the apsidal motion period in close
binaries in terms of stellar masses, relative radii and internal
structure constants of the component stars was given by \citet{russell28}
and later improved by \citet{cowling38}. \citet{chandra33} used polytropic
models to predict internal structure constants for main-sequence stars.
At that time, the large uncertainties of the observational data, as well as
the use of polytropic models with an arbitrary index $n$, were responsible
for the apparently good agreement between observed and predicted values of
$\log k_2$.

By using more realistic stellar models, different authors derived
more elaborated expressions for the apsidal motion period, separating
rotational and tidal contributions to the total apsidal motion rate.
The apsidal motion test was also applied to polytropic models by 
\citet{sterne39}, \citet{brooker55} and, later, to early theoretical
stellar models at the ZAMS, by \citet{schwarz58} and \citet{kushawa57},
both using the old \citet{keller55} opacities. At that time, the theoretical
values of $k_2$ were systematically larger than those obtained by
observations, which led to the interpretation that real stars were more
centrally condensed than predicted by models. This discrepancy persisted
during several decades; more recently, however, \citet{claret02} and
\citet{willems03} have found no systematic effects in the sense that
models are less mass concentrated than real stars, for binaries with accurately
known absolute dimensions.
\citet{jeffery84} and H87 computed
internal structure constants for stars within the main sequence. The former
used \citet{carson76} opacities, while the latter used opacity tables by 
\citet{cox69}. 

In more recent years, the most extensive series of theoretical works in the
literature regarding apsidal motion constants is that one from A.\ Claret and
collaborators, some of which are briefly described here. \citet[hereafter CG89a]{claret89a}
presented a detailed grid of evolutionary stellar models during the hydrogen
burning phases, including apsidal motion constants. \citet{claret91} studied
the effect of the core overshooting and mass loss on the internal density
concentration of main-sequence stars. With more updated input physics,
\citet[hereafter CG92]{claret92} computed stellar models together with internal
together with internal structure constants. Those models showed a general
tendency to be cooler and more centrally concentrated in mass than their
previous computations for the typical masses where apsidal motion is observed.
\citet{claret92,claret93} and \citet{claret95a} were able to reduce the
discrepancies  between theoretical and observed values of $k_2$ at
acceptable levels for systems whose relativistic contributions were small.
\citet{claret99} took into account the effect of rotation on the 
internal structure of stars and found that it strongly depends on the 
distortion of the configuration. \citet{claret02} revised the status of
the apsidal motion test to stellar structure and evolution. They increased
the observational sample by about 50\% in comparison with previous works and took
into account the effects of dynamic tides to determine the contribution of the 
tidal distortion to the predicted apsidal motion rate; they found a good 
agreement between observed and theoretical apsidal motion rates.
The most recent internal structure constants for main-sequence stars are those
contained in \citet[hereafter C04, C05, C06b and C07, respectively]{claret04,
claret05,claret06b,claret07}. They were calculated with new stellar models based
on updated physics computed with different metallicities: (X,\,Z)=(0.70,\,0.02) [C04];
(X,\,Z)=(0.754,\,0.002) and (0.748,\,0.004) [C05]; (X,\,Z)=(0.730,\,0.010)
and (0.739,\,0.007) [C06b]; and (X,\,Z)=(0.64,\,0.04), (0.58,\,0.06) and 
(0.46,\,0.10) [C07].

\vspace{-0.3cm}
\subsection{Perturbations and the apsidal motion}

The longitude of periastron of a binary orbit, $\omega$, 
defines the direction of the line of apsides in the orbital plane. It is
constant, in the orbit of a system consisting of two gravitating bodies,
only if all the three following conditions are valid:
(i) the bodies can be regarded as point masses;
(ii) they move in accordance with Newton law of gravitation ($r^{-2}$); and
(iii) the two bodies form a gravitationally isolated system.
However, if any of these conditions fails, the size, form, and 
spatial position of the orbit will vary. The most readily detectable effect is
a variation in the value of $\omega$ with time that is referred to as rotation
(advance or recession) of the line of apsides. For a more detailed discussion
of this subject, see, for instance, the works of \citet{batten73} or \citet{claret01}.

Several types of perturbations exist and can lead to rotation of apsides, such as
mutual tidal distortion of the components, distortion of the components due to 
axial rotation, relativistic effects, presence of a third body, and recession 
due to a resisting circumbinary medium.
The axial rotation and the mutual tidal forces of close binary systems' components
will deform each other and destroy their spherical symmetry, by means of the 
respective disturbing potentials.  Besides the changes in the stellar structure, 
described in Sect.\,\ref{distortions}, these disturbing potentials produce an observed
change in $\omega$ which is the sum of the effects caused by each component
\citep{batten73}. The total rate of apsidal advance per orbital revolution,
$\dot \omega$, is
\begin{equation}
{\dot \omega \over 2\pi}= {P \over U}=k_{21}c_{21}+k_{22}c_{22},
\label{dotomega}
\end{equation}
where $P$ is the anomalistic orbital period, $U$ is the apsidal
motion period, and
\begin{equation}
c_{2i}=\left[ \left( {\scriptstyle \Omega_i \over \scriptstyle \omega_K} \right) ^2 
\left(\scriptstyle 1+{\scriptstyle M_{3-i} \over \scriptstyle M_i}\right)f(e)
+{\scriptstyle 15M_{3-i} \over \scriptstyle M_i }g(e) \right] \left({\scriptstyle R_i \over\scriptstyle  A} \right)^5,
\label{cconst}
\end{equation}
where $i$=1,2 stands for the primary and the secondary stars, respectively;
$M_i$ and $R_i$ are stellar mass and radius of component $i$; 
$A$ is the semi-major axis; $e$ is the orbital eccentricity;
and the functions $f(e)$ and $g(e)$ are defined as
\begin{equation}
f(e)=(1-e^2)^{-2} ~{\rm and}~
g(e)=\frac{(8+12e^2+e^4)f(e)^{2.5}}{8},  \label{gefunc}
\end{equation}
($\Omega_i/\omega_K$) being the ratio between the actual angular rotational 
velocity of the stars and that corresponding to synchronization with the average
orbital velocity. Equation (\ref{dotomega}) is a special case of
Eq.\,(\ref{isc234}), in which only the second order harmonics are taken into account.
The first term in Eq.\,(\ref{cconst}) represents the contribution to the total apsidal
motion given by rotational distortions and the second term corresponds
to the tidal contributions.

With the exception of the $k_{2i}$, all parameters in Eq.\,(\ref{dotomega}) can be 
determined from photometric and spectroscopic analysis. The empirical
weighted average of the internal structure constants for individual systems
can be given by
\begin{equation}
{\bar k_{2\,\rm obs}}=\frac{1}{c_{21}+c_{22}}\frac{P}{U}=
      \frac{1}{c_{21}+c_{22}}\frac{\dot \omega}{ 2\pi}.
\label{k2obs}
\end{equation}

Equations (\ref{cconst}) and (\ref{k2obs}) show that ${\bar k_{2\,\rm obs}}$ 
depend on our knowledge of the rotation velocities of the component stars.
In most binaries with good absolute dimensions, the rotation velocities of 
the individual components are known through spectroscopic analysis. Since the 
average orbital rotation, or Keplerian velocity, is a function of the orbital 
period, the ratio of rotational velocities in Eq.\,(\ref{cconst}), namely
$\Omega_i/\omega_K$, is well determined in these binaries \citep{claret93}. 
For systems without observational determinations,
the best approach is to assume that the components are synchronized
with the orbital velocity at periastron, where the tidal forces are at maximum.
The rotation velocities are related by \citep{kopal78}
\begin{equation}
\omega _P^2=\frac{(1+e)}{(1-e)^3}\ \omega_K^2,
\label{wperiastron}
\end{equation}
where $\omega _P$ is the angular velocity at periastron, $e$
is the orbital eccentricity, and $\omega_K$ is the Keplerian angular
velocity (Eq.\,\ref{kepler}). \citet{claret93} checked the validity of this 
approximation, achieving a good agreement between the observed and
the predicted rotational velocities by assuming synchronization at periastron
(see their Fig.\,6).

The mean $k_{2i}$ values, obtained through Eq.\,(\ref{k2obs}), can be compared
with those derived from theoretical models (Eq.\,\ref{k2theor}). 
However, the observed mean values of $k_{2\,\rm obs}$ should be first corrected
from non-distortional effects, like relativistic, third body, and
interstellar medium contributions.

\subsection{Internal structure constants for spherically symmetric configurations}
\label{sphericconf}

Internal structure constants ($k_j$) can be approximately computed
based on the simple (and unrealistic) assumption that stars can be
described by spherically symmetric models.
These approximate $k_j$ have been computed for comparison with observed rates
of apsidal motion.
 
The Radau's differential equation \citep{kopal59} is
numerically integrated throughout all the structure with a
4th-order Runge-Kutta method \citep{nrecipes}:
\begin{equation}
\label{radaueq}
r{d\eta _j \over dr} + 6{\rho(r) \over  \bar{\rho}(r)} (\eta_j + 1) 
+ \eta_j (\eta_j - 1) = j(j+1),
\end{equation}
where $\eta_j(0)=j-2$ (j=2,3,4), $\rho(r)$ is the local density at a
distance $r$ from the center, and $\bar{\rho}(r)$ is the mean density
within the inner sphere of radius $r$.
The resulting value of the function $\eta_j(R)$, which satisfies Radau's
equation with $R$ being the radius of the configuration, is used to obtain 
the individual values of $k_j$,
\begin{equation}
k_j={j+1-\eta_j(R) \over 2(j+\eta_j(R))}.
\label{apsidal}
\end{equation}
Because the observed motion of apsides is the sum of the motion produced by both
stars, the quantities $k_{2i}$ ($i$=1,2 for the primary and the secondary
star, respectively) cannot be observationally determined separately. So, the theoretical
counterpart of Eq.\,(\ref{k2obs}) is the weighted mean value
of the individual theoretical apsidal motion constant, 
\begin{equation}
{\bar k_{2,\rm theo}}={c_{21}k_{21,\rm theo}+c_{22}k_{22,\rm theo}\over c_{21}+c_{22}}
,
\label{k2theor}
\end{equation}
where $c_{21}$ and $c_{22}$ are computed through Eq.\,(\ref{cconst}) by using 
absolute dimensions, and
$k_{21, \rm theo}$ and $k_{22, \rm theo}$ are the theoretical apsidal motion 
constants for the primary and the secondary, respectively,
obtained from the stellar models (Eq.\,\ref{apsidal}) for 
the corresponding mass and radius of each component. 
Before performing the apsidal motion comparison, one has to check if
the models are able to reproduce basic stellar parameters, such as 
effective temperatures, and to predict a common age for the two components. A good
discussion of this subject is given by \citet{claret93}. 
In Sect.\,(\ref{comphteoobs}) we show our predictions about apsidal motion
for a chosen binary, comparing them with observational data.

\section{The Kippenhahn \& Thomas formulation}\label{kipthoform}

The KT70 method is a strategy for considering disturbing potentials in
evolutionary stellar models more realistically, so that the distortion
produced by a given disturbing potential is entirely included in the
total potential function.

To clarify how these disturbing effects were taken into account
in the KT70 method, the equations are re-derived here. In this formulation,
the spherically symmetric surfaces, normally used  in standard stellar
models, are replaced by suitable non-spherical  equipotential surfaces
characterized by the total potential $\psi$, the mass $M_{\psi}$ enclosed
by the corresponding equipotential surface whose surface area is $S_{\psi}$
and encloses a volume  $V_{\psi}$, and $r_{\psi}$, the radius of the
topologically equivalent sphere with the same volume $V_{\psi}$, enclosed
by the equipotential surface. 

For any quantity $f$ varying over an equipotential surface, we can define its
mean value as
\begin{equation}
\label{vmedio}
\langle f \rangle = {1 \over S_{\psi}} \int\limits_{\psi_{\rm const.}} fd\sigma ,
\end{equation}
where $S_{\psi} = \int _{\psi_{\rm const.}}d\sigma$, and $d\sigma $ is the
surface element.

The local effective gravity is given by $g = {d\psi \over dn}$,
where $dn$ is the (non-constant) separation between two successive 
equipotentials $\psi $ and $\psi + d\psi $, so that we have
\begin{equation}
\label{gmedio}
\langle g \rangle = {1 \over S_{\psi}} \int\limits_{\psi_{\rm const.}} {d\psi \over dn} d\sigma,
\end{equation}
\begin{equation}
\label{gmedioneg}
\langle g^{-1} \rangle = {1 \over S_{\psi}} \int\limits_{\psi_{\rm const.}} {\left( d\psi \over dn \right) ^{-1}} d\sigma.
\end{equation}
The volume between the surfaces $\psi$ and $\psi$+$d\psi$ is given by
\begin{equation}
\label{dvdpsi}
dV_{\psi }\!\!  = \!\! \int\limits_{\psi_{\rm const.}}\!\!\!\!\!\! dnd\sigma 
        ~  =~d\psi \!\!\!\!\int\limits_{\psi_{\rm const.}}\!\!\! \left( {dn \over d\psi }\right) d\sigma 
	   = d\psi S_{\psi }\langle g^{-1} \rangle ,
\end{equation}
from which we obtain
\begin{equation}
\label{dpsi}
d\psi = {1 \over S_{\psi } \langle g^{-1} \rangle } dV_{\psi } 
      = {1 \over S_{\psi } \langle g^{-1} \rangle } {dM_{\psi } \over \rho (\psi) }{\rm ,}
\end{equation}
and the volume of the topologically equivalent sphere is given by
$V_{\psi} = {4\pi \over 3} r^3 _{\psi}.$

Eq.\,(\ref{dpsi}) can be combined with the general form of the 
hydrostatic equilibrium equation,
\begin{equation}
\label{hideq}
{d P \over d \psi } = - \rho , ~~~~~~~~~~~~ {\mathrm{to~give}}
\end{equation}
\begin{equation}
\label{dpdmpsi}
{dP \over dM_{\psi } } = -{GM_{\psi } \over 4\pi r^4 _{\psi } } f_p ,
\end{equation}
where $f_p$ is given by
\begin{equation}
\label{fpdef}
f_p = {4\pi r^4 _{\psi } \over GM_{\psi } } {1 \over S_{\psi } \langle g^{-1} \rangle }.
\end{equation}
With these corrections, the four stellar structure equations, with
$M_{\psi }$ as the independent variable, become
\begin{ddeqar}
{dP \over dM_{\psi } } & = & - {GM_{\psi } \over 4\pi r^4 _{\psi } }f_p, \nydeqno \\ 
{dr_{\psi } \over dM_{\psi } } & = & {1 \over 4\pi r^2 _{\psi }\rho }, \\
{dL_{\psi } \over dM_{\psi } } & = & \epsilon - T{\partial S \over \partial t }, \\
{dT_{\psi } \over dM_{\psi }} & = & -{GM_{\psi }T \over 4\pi r^4 _{\psi }P}\nabla,
\ \ \nabla = \left\{ \nabla_{\mathrm{rad}}, {f_{\mathrm{t}} \over f_{\mathrm{p}} }\nabla _{\mathrm{conv}} \right\},\ \ \ \ \ \ \ \ \ \ \ \ \    
\arrlabel{starstructeq}
\end{ddeqar}
where $f_p$ is given by Eq.\,(\ref{fpdef}) and 
\begin{equation}
\label{ftdef}
f_t = \left( {4\pi r^2 _{\psi } \over S_{\psi } }\right) ^2 {1 \over \langle g \rangle \langle g^{-1} \rangle }. 
\end{equation}
In the case of isolated and non-rotating stars, $f_p$=$f_t$=1, 
and the original 
stellar structure equations are recovered.
In order to obtain the internal structure of a distorted gas sphere, the 
set of Eqs.\,(\ref{starstructeq}) must be numerically integrated under  
suitable boundary conditions.

This formulation was largely used in the literature mainly due to the easiness
of its implementation in existing evolutionary codes  
\citep[e.g.][]{endal76,law80,pinson88,claret96,mendes}.  

\section{Tidal and/or rotational distortions on the equilibrium structure of stars}
\label{distortions}

We consider the effects
of tidal forces, as well as the combined effects of tidal forces and rotation,
on the stellar tructure and implemented such effects in the {\tt ATON} code.
\cite{mendesphd} had already introduced the effects
of rotation alone in it. 
The tidal and rotational effects in stellar models were implemented according to
the treatment derived by KT70 and modified by \citet{endal76}, who in turn used
a more refined function to take into account terms related to the distortion
of the stars, namely the Clairaut-Legendre expansion for the gravitational
potential of a self-gravitating body \citep{kopal59}. 
The calculations are done within the framework of static tides; in the
case of dynamic tides, a more refined treatment is required
\citep[see][]{claret02,willems03}.

In our model, tides and axial rotation are the two physical causes of
the stellar configuration deviating from a spherical form. 
The centrifugal pseudo-potential terms arising from
the orbital motion around the center of mass of the system do contribute,
also, for the departure from sphericity, but we do not consider these
terms in the present work, yet. These terms will be included in future works.
We treat three different situations: 1) rotation acting alone, 
2) tidal forces acting alone and 3) a combination of both effects
distorting the star. The special case of the evolution of 
one of the components in a binary system is considered, assuming that
(i) it rotates about an axis perpendicular to the
orbital plane; (ii) from the orbital point of view, the system is 
considered to be at rest,i.e., the stars do not revolve around each other;
and, particularly
in this work, (iii) the ratio of the mutual separation to the radius of the 
evolving (distorted) star is assumed to be constant. The value of
this ratio is an input parameter that depends on the system to be
reproduced; in close binaries, the separation of the stars is often
less than 10 times their radii \citep{hilditch01}.
The equipotential surfaces can be written as an expansion of the tesseral 
harmonics\footnote{Tesseral harmonics are 
spherical harmonics of the form $\mathrm{\cos (m\phi)P_l^m(\cos\theta)}$ and 
$\mathrm{\sin (m\phi)P_l^m(\cos\theta)}$, for $\mathrm{m \neq l}$. These harmonics 
are so named because the curves on which they vanish are 
$l-m$ parallels of latitude and $2m$ meridians, which divide the 
surface of a sphere into quadrangles whose angles are right 
angles.} $Y^i_j$, 
\begin{equation}
r(r_0,\theta,\phi)=r_0\left[1+\sum_{i,j}Y^i_j(r_0,\theta,\phi)\right],
\label{genraleqsup}
\end{equation}
where $r_0$ is the mean radius of the corresponding equipotential surface.
The tesseral harmonics can be written as
\begin{equation}
Y^i_j(r_0,\theta,\phi)=K(r_0)P^i_j(\theta,\phi),
\label{yfact}
\end{equation}
where $K(r_0)$ is the radial part of the tesseral harmonics and 
its angular part, $\mathrm{P^i_j}$, being functions associated
with Legendre polynomials.

\subsection{Rotational distortion}\label{rotsect}

Rotation alone would render the star a rotational spheroid flattened
at the poles. In what follows we are considering
conservative rotation, which means that the centrifugal acceleration
can be derived from a potential,
\begin{equation}
    \Omega^2 s\, \mathrm{e}_s = - \nabla V_{\rm rot},
    \label{eq:conserv}
\end{equation}
where $s=r\sin \theta$ is the perpendicular distance to the rotation
axis of the star. A necessary and sufficient condition for the
conservative case is that $\Omega=\Omega(s)$ \citep{kip:94}, which
means rotation is constant on cylinders. Rigid body rotation is
obviously a special case of conservative rotation.

Following \citet{kopal59} and later \citet{endal76}, the
total potential is divided in three parts according to Eqs.\,(\ref{termspotrot}),
where $\psi _s$ is the spherically symmetric part of the gravitational potential,
$\psi _r$ is the cylindrically symmetric potential due to rotation, and $\psi _d^{\rm (rot)}$
is the cylindrically symmetric part of the gravitational potential due to
the distortion of the figure of the star caused by rotation. If the coordinates of
the point $P$ are the radius $r$ and the polar angle $\theta$ 
(measured from the rotational axis), the  
components of the potential at $P$ can be written as
\begin{deqarr}
\psi_s & = & {GM_{\psi} \over r}, \\
\psi_r & = & {1 \over 2} \Omega ^2 \sin ^2 \theta, \label{rotpot} \\
\psi_d^{\rm (rot)} & = & \sum_{j=2}^{\infty}{{4\pi G} \over {(2j+1)r^{j+1}}}
    \int\limits^{r_0}_0 \rho {\partial \over {\partial r_0^{\prime}}}
    (r_0^{\prime j+3} Y^i_j) dr_0^{\prime}.
\arrlabel{termspotrot}
\end{deqarr}
In Eqs.\,(\ref{termspotrot}), $M_\psi$ and $\Omega$ are respectively the mass and the
angular velocity of the rotating star, while $r_0$, by virtue
of the cylindrically symmetric rotational potential $\psi_r$, corresponds to the radius
of the equipotential surface (as given by Eq. \ref{genraleqsup})
at the angle $\theta _0$ defined such that $P_2(\cos \theta _0)$=0, $P_2$
being the second-order Legendre polynomial.
Here, the shape of rotating configurations is described according to 
the expansion given in Eq.\,(\ref{genraleqsup}), with $Y_j^i$ given by
\begin{equation}
  Y_j^i(r_0) = c_{i,j} {{2j + 1} \over {j + \eta_j(r_0)}} {r_0^{j+1} \over {GM_\psi}}
               P^i_j(\theta,\phi).
\end{equation}

By limiting ourselves to first-order theory, the disturbing potential associated
to rotation (Eq. \ref{rotpot}) is such that it will invoke a single non-zero $Y^i_j$ term in Eq.
(\ref{genraleqsup}), namely that one corresponding to $i\!=\!0,\,j\!=\!2$ \citep{kopal59,kopal60}:
\begin{equation}
 c\,_{0,2} = - {1 \over 3} \Omega^2.
\end{equation}
So, in the case of rotational distortions acting alone, the equipotential surface now becomes
\begin{equation}
\label{othereqsurrot}
r(r_0,\theta) = r_0 \Big[ 1 +Y_{\rm rot} \Big].
\end{equation}
in which we replaced $Y_2$ by $Y_{\rm rot}$ for better clarity, and
\begin{equation}
\label{hartessrot}
Y_{\rm rot} =  - {\Omega ^2 r_0 ^3 \over 3GM_{\psi}} {5 \over 2 + \eta _2 (r_0)} P_2 (\cos \theta ),
\end{equation}
where we have abbreviated, as it is customary,  
\begin{equation}
\label{etarot1}
\eta_2 = {r_0 \over Y_{\rm rot}} {\partial Y_{\rm rot} \over \partial r_0}.
\end{equation}
In short, $Y_{\rm rot}$ is a measure of the deviation
from sphericity caused by rotation.
The quantity $\eta_2$ is of particular interest to our study, because
the theoretical apsidal motion constant $k_2$ can be derived from it
(see Sect.\,\ref{apsmotion}). The evaluation of $\eta_2$ can be done by numerically
integrating the Radau's equation:
\begin{equation}
\label{radaueqrot}
r_0{d\eta_2 \over dr_0} + 6{\rho(r_0) \over \bar{\rho}(r_0)} (\eta_2 + 1) + \eta_2 (\eta_2 - 1) = 6.
\end{equation}
This equation is slightly different from Eq.\,(\ref{radaueq}). Here we use $j=2$, and the
spherical radius $r$ was replaced by $r_0$, the mean radius of the distorted configuration.

By virtue of the definition of $Y_{\rm rot}\equiv Y_2$ as given by Eq.
(\ref{hartessrot}), the expression for $\psi_d^{\rm (rot)}$ simplifies to
\begin{equation}
 \psi_d^{\rm (rot)} = - {4\pi \over 3r^3 } P_2(\cos \theta ) \int\limits^{r_0}_0 \rho
          {{r^{\prime}}^7 _0 \over M_{\psi }} \Omega ^2 {5+\eta _2 \over 2+\eta _2 }
          dr^{\prime}_0
\end{equation} so that the total potential can be written as
\begin{eqnarray}
 \psi & = &\psi _s + \psi _r + \psi _d^{\rm (rot)} \nonumber \\
      & = & {GM_{\psi} \over r} + {1 \over 2} \Omega ^2 \sin ^2 \theta + \nonumber \\ 
      &   & - {4\pi \over 3r^3 } P_2(\cos \theta ) \int\limits^{r_0}_0 \rho
            {{r^{\prime}}^7 _0 \over M_{\psi }} \Omega ^2 {5+\eta _2 \over 2+\eta _2 }
            dr^{\prime}_0. \label{totrotpot}
\end{eqnarray}

By defining the radial part of the axisymmetric tesseral harmonic $Y_{\rm rot}$ as 
\begin{equation}
\label{yaconst}
A(r_0)={\Omega ^2 r_0 ^3 \over 3GM_{\psi}} {5 \over 2+\eta _2 },
\end{equation}
the radius of the equipotential surface, Eq.\,(\ref{othereqsurrot}), can be rewritten as
\begin{equation}
\label{neweqsurrot}
r(r_0,\theta) = r_0 \Big[1-A(r_0)P_2 (\cos \theta ) \Big].
\end{equation}
To relate $r_0$ to $r_{\psi}$, we evaluate the volume integral from
$r$=0 to $r(r_0,\theta)$ to obtain
\begin{equation}
\label{newvpsirot}
V_{\psi} = {4 \pi r_0 ^3 \over 3} \Biggl[ 1 + {3 \over 5}A^2  - {2 \over 35}A^3 \Biggr].
\end{equation}
For simplicity, the arguments of the term $A(r_0)$
were suppressed. From the previous equation, $r_{\psi}$ is then obtained as
\begin{equation}
\label{rpsirot}
r_{\psi}  =  r_0 \Biggl[  1 + {3 \over 5}A^2 - {2 \over 35}A^3 \Biggr]^{1/3}.
\end{equation}
Usually, $r_{\psi }$ is known and so $r_0 $ can be calculated through
Eq.\,(\ref{rpsirot}) by means of an iterative procedure.

Since the local effective gravity is given by
\begin{equation}
\label{locgravrot}
g  =  {\partial \psi \over \partial n } = \left[ \left(
      {\partial \psi \over \partial r }\right)^2 + \left(
      {1 \over r }{\partial \psi \over \partial \theta }\right) ^2  \right] ^{1/2},
\end{equation}
$g$ can be found by differentiation of Eq.\,(\ref{totrotpot}). The
integral in Eq.\,(\ref{totrotpot}) and their derivatives must be evaluated
numerically. Once the values of $\langle g \rangle $ and $\langle g^{-1} \rangle$
are known for a set of points on an equipotential surface,
$S_{\psi }\langle g \rangle $ and $S_{\psi } \langle g^{-1} \rangle$
can be found, respectively, from Eqs.\,(\ref{gmedio}) and (\ref{gmedioneg})
by numerically integrating over $\theta$.
%
\subsection{Tidal distortion}\label{tidsect} 
Tidal distortion acting alone would tend to elongate the star in the
direction of the other component.
Similarly to Eqs.\,(\ref{termspotrot}), the total potential is divided
in three parts $\psi_s$, $\psi _t$, and $\psi_d^{\rm (tid)}$, where
$\psi_s$ is again the spherically symmetric part of the gravitational,
potential, $\psi _t$ is the non-symmetric potential due to tidal forces,
and $\psi_d^{\rm (tid)}$ is the non-symmetric part of the gravitational potential due to
distortion of the star caused by the presence of the companion.
If the coordinates of the point $P$ are the radius $r$, the polar
angle $\theta$, and the azimuthal angle $\phi$, the components of the
potential at $P$ can be written as:
\begin{deqarr}
\psi_s & = & {GM_{\psi} \over r}, \\
\psi_t & = & {GM_2 \over R} \Bigg[ \sum_{j=2}^\infty \left( {r_0 \over R} \right)^j P_j
(\lambda ) \Bigg], \label{labtid} \\
\psi_d^{\rm (tid)} & = & \sum_{j=2}^{\infty}{{4\pi G} \over {(2j+1)r^{j+1}}}\!\int\limits^{r_0}_0\!\rho
{\partial \over {\partial r_0^{\prime}}} (r_0^{\prime j+3} Y^i_j) dr_0^{\prime} \label{labpsid}.
\arrlabel{termspottid}
\end{deqarr}
In Eqs.\,(\ref{termspottid}),
$M_\psi$ and $M_2$ are respectively the masses of the primary and the
disturbing star; $R$ is the mutual separation between the centers
of mass of the two stars; $\lambda$=$\cos \phi \sin \theta$;
$r_0$ is the radius of the equipotential surface at the angles
$(\theta_0,\phi_0)$, defined such that 
\begin{equation}
  \mathrm{B(r_0)P_2(\lambda_0)+C(r_0)P_3(\lambda_0)+D(r_0)P_4(\lambda_0)= 0},
\end{equation}
where $\mathrm{\lambda_0=\cos\phi_0 \sin\theta_0}$ and the terms 
$\mathrm{B(r_0)}$, $\mathrm{C(r_0)}$ and $\mathrm{D(r_0)}$ are defined in 
Eqs.\,(\ref{ybcdconst}); and $\mathrm{P_j}$ are the $\mathrm{j}$th-order
Legendre polynomials.

By virtue of the disturbing tidal potential given by Eq. (\ref{labtid}), and again
restricting ourselves to first-order quantities only, it follows that the only non-zero
$Y_j$ terms in Eq. (\ref{genraleqsup}) consistent with the equilibrium theory of tides are
those for which $i\!=\!0,\, j\!=\!2,3,4$ (the reader is referred to the works of
\citealt{kopal59, kopal60} for full details):
\begin{equation}
c\,_{0,j} = G {M_2 \over R^{j+1}} \ \ \ (j=2, 3, 4),
\end{equation}
from which we have
\begin{equation}
\label{hartesstid}
Y_j = {M_2 \over M_{\psi}} \  {2j +1 \over j + \eta _j (r_0)}\  
\left( {r_0 \over R} \right) ^{j+1} P_j (\lambda), \ \ \ (j=2, 3, 4),
\end{equation}
where, as before,
\begin{equation}
\label{etatid}
\eta _j = {r_0 \over Y_j} {\partial Y_j \over \partial r_0};
\end{equation}
the equipotential surfaces are then described by
\begin{equation}
\label{othereqsurtid}
r(r_0,\theta ,\phi) = r_0 \bigg[ 1 + \sum^4_{j=2} Y_j \bigg].
\end{equation}
So, for tidal forces acting alone, the $Y_j$ are a measure of
the deviation from sphericity due to those forces.

As in Sect.\,\ref{rotsect}, the apsidal motion constants can be derived
from $\eta_j$
by using our theoretical stellar models (see Sect.\,\ref{apsmotion}).
The evaluation of the quantity above can be done by, again, numerically
integrating the Radau's equation
\begin{equation}
\label{radaueqtid}
r_0{d\eta _j \over dr_0} + 6{\rho(r_0) \over \bar{\rho}(r_0)} (\eta
_j + 1) + \eta _j (\eta _j - 1) = j(j+1),
\end{equation}
for each $j=2,3,4$.

Substituting $Y_j$ as given by Eq. (\ref{hartesstid}) in Eq.
(\ref{labpsid}) and carrying on the partial derivative in the latter, we get
\begin{equation}
\psi_d^{\rm (tid)} = 4\pi GM_2 \sum\limits_{j=2}^4 {P_j(\lambda ) \over (rR)^{j+1}}
         \int\limits^{r_0}_0 \rho {{r^{\prime}}^{2j+3}_0 \over M_{\psi }} 
         {j+3+\eta_j \over j+\eta_j } dr^{\prime}_0.
\end{equation}
The total potential is then
\begin{eqnarray}
 \psi & = &\psi _s + \psi _t + \psi _r^{\rm (tid)} \nonumber \\
      & = & {GM_{\psi} \over r}  +{GM_2 \over R} \Bigg[ \sum_{j=2}^\infty
          \left( {r_0 \over R} \right)^j P_j (\lambda ) \Bigg] +  \nonumber \\  
      &   & +\, 4\pi GM_2 \sum\limits_{j=2}^4 {P_j(\lambda ) \over (rR)^{j+1}}
          \int\limits^{r_0}_0 \rho {{r^{\prime}}^{2j+3}_0 \over M_{\psi }} 
          {j+3+\eta_j \over j+\eta_j } dr^{\prime}_0 \label{tottidpot}.
\end{eqnarray}

By defining the radial parts of the non-symmetric tesseral harmonics $Y_j$
in Eq. (\ref{hartesstid}) as
\begin{ddeqar}
B(r_0)& = & {M_2 \over M_{\psi}} \left( {r_0 \over R } \right) ^3 {5 \over 2+\eta _2 },
\nydeqno \\   
C(r_0)&=&{M_2 \over M_{\psi}} \left( {r_0 \over R } \right) ^4 {7 \over 3+\eta _3 }, \\
D(r_0)&=&{M_2 \over M_{\psi}} \left( {r_0 \over R } \right) ^5 {9 \over 4+\eta _4 },
\arrlabel{ybcdconst}
\end{ddeqar}
we can describe the equipotential surfaces, Eq.\,(\ref{othereqsurtid}), as
\begin{eqnarray}
\lefteqn{r(r_0,\theta,\phi) = } \nonumber \\
  & & r_0 \bigl[1 + B(r_0)P_2 (\lambda ) + C(r_0)P_3 (\lambda )
                  + D(r_0)P_4 (\lambda ) \bigr].
\label{neweqsurtid}
\end{eqnarray}
We evaluate the volume integral from $r$=0 to $r(r_0,\theta,\phi)$ given 
by Eq.\,(\ref{neweqsurtid}) and obtain
\begin{eqnarray}
\label{newvpsitid}
V_{\psi}\!\!&=\!\displaystyle\frac{4 \pi r_0 ^3}{3} \Biggl[\!\! &1 + \frac{3 B^2}{5} + \frac{3 C^2}{7} 
+ \frac{2 B^3}{35} + \frac{6B^2D}{35} + \frac{D^2}{3} \nonumber \\
& & + \frac{4 BC^2}{35} + \frac{20 BD^2}{231} + \frac{6 C^2D}{77} 
+ \frac{18 D^3}{1001} ~\Biggr],
\end{eqnarray}
where the arguments of the terms $B(r_0)$,
$C(r_0)$ and $D(r_0)$ were omitted. From Eq.\,(\ref{newvpsitid}),
$r_0$ and $r_{\psi}$ are related by
\begin{eqnarray}
\label{rpsitid}
r_{\psi}& =  r_0 \Biggl[ & 1 + \frac{3 B^2}{5} + \frac{3C^2}{7} + \frac{2 B^3}{35}
                           + \frac{6B^2D}{35} + \frac{4 BC^2}{35}\nonumber \\
    &                    & + \frac{20 BD^2}{231} + \frac{6 C^2D}{77} 
                           + \frac{18 D^3}{1001} + \frac{D^2}{3}~~\Biggr]^{1\over 3}.
\end{eqnarray}
Similarly to the case of pure rotation, $r_0 $ can be calculated 
from the value of $r_\psi$ through Eq.\,(\ref{rpsitid}), by means of an iterative
procedure.

The local effective gravity is now given by
\begin{equation}
\label{locgravtid}
g  =  {\partial \psi \over \partial n } = \left[ \left(
      {\partial \psi \over \partial r }\right) ^2 + \left(
      {1 \over r }{\partial \psi \over \partial \theta }\right)^2 +
      \left( {1 \over r \sin \theta }{\partial \psi \over \partial \phi }
      \right)^2 \right] ^{1\over 2}
\hspace{-0.2cm},
\end{equation}
and it can be obtained by differentiation of Eq.\,(\ref{tottidpot}).
As before, integrals and derivatives must be evaluated numerically. 

Finally, by integrating over $\theta$ and $\phi$, we find 
$S_{\psi }\langle g \rangle $ and $S_{\psi } \langle g^{-1} \rangle$
from Eqs.\,(\ref{gmedio}) and (\ref{gmedioneg}), respectively, once 
$\langle g \rangle $ and $\langle g^{-1} \rangle $
are known for a set of points on an equipotential surface.
 
\subsection{Combined effects of rotation and tides}\label{tidalrotsect}

According to \cite{kopal60,kopal74}, the combined effects of rotation and tides
give rise to a total disturbing potential that can be expressed as the sum of the distortion
terms corresponding to pure rotation or tides and of a number of interaction terms. The
latter, however, are in general second-order terms; so, in the present work, we neglect those
interaction terms since we are dealing with first-order theory only.

The total potential for the combined case is then composed
by four parts, which represent the joint contributions from first-order
rotational and tidal effects as respectively given by Eqs.\,(\ref{termspotrot}) and (\ref{termspottid}):
$\psi _s$, the spherically symmetric part of the gravitational potential;
$\psi _r$, the cylindrically symmetric potential due to rotation; 
$\psi _t$, the non-symmetric potential due to tidal forces;
and $\psi _d\equiv\psi_d^{\rm (rot)}+\psi_d^{\rm (tid)}$,
the non-symmetric part of the gravitational potential due the to distortion
of the star considering both effects.
Again, we do not take into account
the pseudo-potential centrifugal terms due to the
orbital motion in this approximation.
So, the total potential for the combined case (to first-order approximation)
at $P(r,\theta,\phi)$ is
\begin{eqnarray}
  \psi & = & \psi _s + \psi _r + \psi_t + \psi _d \nonumber \\
       & = & {GM_{\psi} \over r} + {1 \over 2} \Omega ^2 \sin ^2 \theta + 
             {GM_2 \over R}\Bigg[1+ \sum_{j=2}^4 \left( {r_0 \over R} \right)^j P_j
             (\lambda) \Bigg]  \nonumber \\ 
       &   & - {4\pi \over 3r^3 } P_2(\cos \theta ) \int\limits^{r_0}_0 \rho
             {{r^{\prime}}^7 _0 \over M_{\psi }} \Omega ^2 {5+\eta _2 \over 2+\eta _2}
             dr^{\prime}_0  \nonumber \\
       &   & + 4\pi GM_2 \sum_{j=2}^4 {P_j(\lambda ) \over (rR)^{j+1}}
             \int\limits^{r_0}_0 \rho {{r^{\prime}}^{2j+3}_0 \over M_{\psi}}
             {j+3+\eta_j \over j+\eta_j } dr^{\prime}_0. \label{totpot}
\end{eqnarray}
In Eq.\,(\ref{totpot}), $M_\psi$, $M_2$, $\Omega$, $R$, and $\lambda$ retain their
meanings as previously defined in Sects.\,(\ref{rotsect}) and (\ref{tidsect}),
while $r_0$ is now the radius of the equipotential surface at the angles
$(\theta_0,\phi_0)$ defined such that
\begin{eqnarray}
   \lefteqn{-A(r_0)P_2(\cos \theta _0)+B(r_0)P_2(\lambda_0)+C(r_0)P_3(\lambda_0)} \nonumber \\
   & & + D(r_0)P_4(\lambda_0)= 0,
\end{eqnarray}
where A($r_0$) is given by Eq.\,(\ref{yaconst}), and B($r_0$), C($r_0$), and D($r_0$) are 
defined in Eqs.\,(\ref{ybcdconst}).

Since, as previously mentioned, the distortions arising from
both rotation and tidal effects are simply additive (to first- and even
second-order accuracy; \citealt{kopal60,kopal74,kopal89}), the external equipotential
surface of the distorted star can be described by
\begin{equation}
\label{othereqsur}
r(r_0,\theta ,\phi) = r_0 \bigg[ 1 +Y_{\rm rot} + \sum^4_2 Y_j \bigg],
\end{equation}
where  $Y_{\rm rot}$ is given by Eq.\,(\ref{hartessrot}) and the $Y_j$ are given
by Eq.\,(\ref{hartesstid}).
As in Sects.\,(\ref{rotsect}) and (\ref{tidsect}), the evaluation of  
$\eta_j$ can be done by numerically integrating Radau's equation (Eq.\,\ref{radaueqtid}).
In this way, by using an adequate iterative procedure, we can obtain a value for
$\eta_2$ \textrm{that} reflects the combined effects of rotation and tides, and $\eta_3$ and $\eta_4$
values that obviously relate only to tidal effects.    
 
If we define the radial parts of the tesseral harmonics, 
$A(r_0)$, $B(r_0)$, $C(r_0)$ and $D(r_0)$, according to Eqs.\,(\ref{yaconst}) and
(\ref{ybcdconst}), Eq.\,(\ref{othereqsur}) can be rewritten as
\begin{eqnarray}
\label{neweqsur}
r(r_0,\theta ,\phi) & = r_0 \Big[&1-A(r_0)P_2 (\cos \theta ) + 
                          B(r_0)P_2 (\lambda ) \nonumber \\ 
                    &   & + C(r_0)P_3 (\lambda ) + D(r_0)P_4 (\lambda ) ~~ \Big].
\end{eqnarray}
By integrating Eq.\,(\ref{neweqsur}) from $r$=0 to $r(r_0,\theta,\phi)$, we obtain
\begin{eqnarray}
V_{\psi}& = \displaystyle {4 \pi r_0 ^3 \over 3} \biggl[& 1 + \frac{3 A^2}{5} 
+ \frac{3 AB}{5} + \frac{3  B^2}{5} + \frac{3 C^2}{7} - \frac{2 A^3}{35} 
\nonumber \\ 
&&- \frac{3 A^2B}{35} + \frac{9 A^2D}{140} + \frac{3 AB^2}{35} + \frac{2 AC^2}{35} 
+ \frac{2 B^3}{35} \nonumber \\ 
&&+ \frac{10 AD^2}{231} + \frac{6 ABD}{35} + \frac{6 B^2D}{35} + \frac{4 BC^2}{35}
\nonumber \\ 
&&+ \frac{6 C^2D}{77} + \frac{20 BD^2}{231} + \frac{18 D^3}{1001} 
+ \frac{D^2}{3} ~~\biggr].\,~~~ 
\label{newvpsi}
\end{eqnarray}
For simplicity, $A(r_0)$, $B(r_0)$, $C(r_0)$, $D(r_0)$ appear with no arguments. 
From Eq.\,(\ref{newvpsi}), $r_0$ and $r_{\psi}$ are related as
\begin{eqnarray}
\label{rpsi}
r_{\psi}& =  r_0 \biggl[& 1 + \frac{3 A^2}{5} 
+ \frac{3 AB}{5} + \frac{3  B^2}{5} + \frac{3 C^2}{7} - \frac{2 A^3}{35} 
\nonumber \\ 
&&- \frac{3 A^2B}{35} + \frac{9 A^2D}{140} + \frac{3 AB^2}{35} + \frac{2 AC^2}{35} 
+ \frac{2 B^3}{35} \nonumber \\ 
&&- \frac{3 A^2B}{35} + \frac{9 A^2D}{140} + \frac{3 AB^2}{35} + \frac{2 AC^2}{35} 
+ \frac{2 B^3}{35} \nonumber \\ 
&&+ \frac{6 C^2D}{77} + \frac{20 BD^2}{231} + \frac{18 D^3}{1001} 
+ \frac{D^2}{3}~~ \biggr]^{1\over3}.\,~~~ 
\end{eqnarray}
As before, $r_0 $ can be calculated through 
Eq.\,(\ref{rpsi}) by means of an iterative procedure.

The local effective gravity is given by Eq.\,(\ref{locgravtid}),
$g$ can be numerically found by differentiation of Eq.\,(\ref{totpot}).
With the values of
$\langle g \rangle $ and $\langle g^{-1} \rangle $ known for a set of 
points on an equipotential surface, 
$S_{\psi }\langle g \rangle $ and $S_{\psi } \langle g^{-1} \rangle$
can be found, respectively, from Eqs.\,(\ref{gmedio}) and (\ref{gmedioneg}) 
by numerically integrating over $\theta$ and $\phi$.

\subsection{Rotational inertia} \label{rotinertia}

Rotational inertia (or moment of inertia) is an important tool for studying
tidal evolution theories, since it is
required to predict the circularization and synchronization time scales 
\citep{zahn77}.

\citet{motz52} computed values of rotational inertia by using the 
existing (and unrealistic) models at that time. 
\citet{rucinski88} obtained values of the 
radii of gyration for polytropic models of low-mass stars at the ZAMS.
The radius of gyration of a body is the distance between a given axis 
of this body\footnote{In a rotating body, the rotation axis is considered. If
no axis is specified, the centroidal axis, which is the line joining
the centroid of each cross section along the length of an axial
member such as truss diagonal, is assumed.} and its center 
of gyration\footnote{The center of gyration of a body is defined as that
point at which the whole mass might be concentrated (theoretically) without 
altering the body's rotational inertia. In other words, this is the center
about which the body can rotate without moving linearly or vibrating.}, 
and is defined by
\begin{equation}
\beta =\sqrt{{I \over MR^2}},
\label{beta}
\end{equation}
where $I$ is the rotational inertia of the star, and $M$ and $R$ are the stellar mass 
and radius, respectively.
\citet{claret89b} presented radii of gyration calculations
for more massive stars during the hydrogen burning phases. They used standard models,
in which the stars are described by spherically symmetric configurations. 
More recent computations of radii of gyration were provided by 
C04, C05, C06b, and C07.

As a consequence of the distortions introduced
by rotation and tides, the rotational inertia of the star is
changed from that one corresponding to spherical symmetry. \citet{law80}
derived the rotational inertia of a rotationally distorted mass shell, to first-order
accuracy, as
\begin{equation}
\Delta I={2\over 3}dm_{\psi}r_{\psi}^2\left({r_0 \over r_{\psi}}\right)^4
\biggl[1+{3\over 20}\sum_{i=1}^5\alpha_iA^i(i\eta_2+5)\biggr],
\label{monIrota}
\end{equation}
where 
\begin{equation}
\alpha_i={5\over i!(5-i)!}\int_0^{\pi}P_2^i(\cos\theta)\sin^3\theta d\theta.
\label{alphai}
\end{equation}

In what follows, we present a new expression for the rotational inertia
taking into account both rotational and tidal distortions, again considering
only first-order effects. We start by considering a very thin mass shell,
for which the rotational inertia is given by
\begin{equation}
\Delta I = \int R^2 dm.
\label{momI1}
\end{equation}
By using spherical coordinates, so that $R=r\sin\theta$ and
$dm=\rho r^2\sin\theta dr d\theta d\phi$, and assuming that $\rho$ is  
constant in such a thin mass shell (which will be later justified as we get to the
limiting differential case), the previous expression becomes
\begin{eqnarray}
  \Delta I & = & \rho \int\limits_0^{2\pi} \int\limits_0^{\pi}
               \int\limits_{r_1}^{r_2} r^4 \sin^3\theta dr d\theta d\phi \nonumber \\
           & = & {\rho \over 5}  \int\limits_0^{2\pi}
               \int\limits_0^{\pi}(r_2^5-r_1^5)\sin^3\theta d\theta d\phi{\rm {\,,}}
\label{monI2}
\end{eqnarray}
where $r_1$ and $r_2$ are respectively the inner and outer radius of the mass
shell. For a tidally and rotationally distorted mass shell, $r_1$ and $r_2$
correspond to the radius of the equipotential surfaces of the distorted configuration
as given by Eq. (\ref{neweqsur}), so that

\begin{eqnarray}
r_i & = r_{0i}\bigl[ &1-A(r_{0i})P_2(\cos\theta)+B(r_{0i})P_2(\lambda) + \nonumber \\
    &   & +\, C(r_{0i})P_3(\lambda)+D(r_{0i})P_4(\lambda)~~\bigr],
\label{r1r2}
\end{eqnarray}
where $i$=1,2.

Since the terms $B(r_{0i})$, $C(r_{0i})$ and $D(r_{0i})$, in Eq.\ (\ref{r1r2}), are
proportional to $(r_{0i}/R)^j$ with $j=2,3,4$ respectively, and that, furthermore,
they would be raised to the fifth power when substituted in Eq. (\ref{monI2}),
it follows that $B(r_{0i})$ would dominate over $C(r_{0i})$ and $D(r_{0i})$ in the
latter equation, as $r_{0i}/R\!<\!1$. This, in short, means that terms of order
$j$ higher than 2 do not contribute significantly to the total departure from
the spherical symmetry, as already noted by \cite{claret02}. So, as a further
approximation, we drop the terms corresponding to $j=3,4$ in Eq. (\ref{r1r2})
for the sole purpose of computing the rotational inertia of a mass shell distorted
by tides and rotation. This approximation allows us to reduce from 252 to 42
the terms that result from the full development of Eq.\ (\ref{monI2}) when $r_1$ and
$r_2$ are substituted for those given by Eq.\ (\ref{r1r2}). In this way, $\Delta I$
becomes, after some algebra,
\begin{eqnarray}
\label{monI3}
\Delta I& = &\displaystyle {\rho \over 5}\int _0 ^{2\pi}\int _0 ^{\pi}\biggl\{
\sum_{i=1}^2 (-1)^i r_{0i}^5\Big[1-A(r_{0i})P_2(\cos\theta) + \nonumber \\ 
&&+B(r_{0i})P_2(\lambda)\Big]^5 \biggl\} \sin^3\theta d\theta d\phi.
\end{eqnarray}
The fifth power term within  brackets in Eq. (\ref{monI3})
can be expanded by using the multinomial theorem
\begin{eqnarray}
   \lefteqn{(x_1+x_2+...+x_p)^n=} \nonumber \\
   & &  \sum_{0\leq a_1,a_2,...,a_p\leq n \atop a_1+a_2+...+a_p=n}
        {n \choose a_1,a_2,...,a_p}\,x_1^{a_1}\,x_2^{a_2}\,\cdots\,x_p^{a_p},
\label{multinon}
\end{eqnarray}

yielding
\begin{eqnarray}
\label{monI5}
\Delta I\! = \!{\rho \over 5} \!\!\!
\sum_{0\leq a_1,a_2,a_3\leq 5 \atop a_1+a_2+a_3=5} \!\!\!Q_a
\Biggl[\sum_{i=1}^2 (-1)^{i+1}r_{0i}^5A^{a_2}(r_{0i})B^{a_3}(r_{0i}) \Biggl],
\end{eqnarray}
where
\begin{equation}
Q_a\!\!=\!\!\Biggl( {5! \over a_1! a_2! a_3!} \Biggl) 
\int_0^{2\pi}\!\!\! \int_0^{\pi} P_2^{a_2}(\cos\theta)P_2^{a_3}(\lambda)
\sin^3\theta d\theta d\phi.
\label{ka}
\end{equation}
For the limiting case of a thin mass shell, we have
\begin{eqnarray}
r_{01}&=&r_0, \nonumber \\
r_{02}&=&r_{01}+dr_{01}=r_0+dr_0, \nonumber \\
A(r_{01})&=&A,\nonumber \\
B(r_{01})&=&B, \nonumber \\
A(r_{02})&=&A(r_{01})+dA(r_{01})=A+dA ~~~~ {\rm and } \nonumber \\
B(r_{02})&=&B(r_{01})+dB(r_{01})=B+dB,\nonumber
\end{eqnarray}
so that the terms within brackets in Eq. (\ref{monI5})
can be rewritten as
\begin{eqnarray}
\label{brack}
\lefteqn{\biggl[ -r_{02}^5A^{a_2}(r_{02})B^{a_3}(r_{02})+
r_{01}^5A^{a_2}(r_{01})B^{a_3}(r_{01}) \biggl] =} \nonumber \\
&&\biggl[ - (r_0+dr_0)^5(A+dA)^{a_2} (B+dB)^{a_3}+
r_0^5A^{a_2}B^{a_3}\biggl]\,.
\label{eqbracket}
\end{eqnarray}
For a thin mass shell one has $r_0\!\gg\!dr_0$ and
consequently $A\!\gg\!dA$ and $B\!\gg\!dB$, so the right side of Eq. (\ref{eqbracket})
can be further reduced to
\begin{displaymath}
  -r_0^4 dr_0 A^{a_2} B^{a_3} \biggl[a_2 \eta_2^{\rm (rot)} + a_3 \eta_2^{\rm (tid)} +5 \biggr],
\end{displaymath}
where $\eta_2^{\rm (rot)}$ and $\eta_2^{\rm (tid)}$ stand for the $\eta_2$ values that
would result from pure rotation or pure tidal effects, respectively.

By using this latter result and remembering that
$dm_{\psi}=4\pi\rho r^2_{\psi}dr_{\psi}$,
Eq. (\ref{monI5}) can be brought to the form
\begin{eqnarray}
\lefteqn{\Delta I= -{1 \over 20\pi}dm_{\psi}
{r_0^4 \over r_{\psi}^2} {dr_0 \over dr_{\psi}} \: \times} \nonumber \\
&&~~~~~~\sum_{0\leq a_1,a_2,a_3\leq 5 \atop a_1+a_2+a_3=5} Q_a
A^{a_2}B^{a_3}\biggl[a_2\eta_2^{\rm (rot)}+a_3\eta_2^{\rm (tid)}+5\biggl],
\label{monI7}
\end{eqnarray}
where $Q_a$ is given by Eq.\,(\ref{ka}).

In the special situation in which only tidal forces are present,
the rotational inertia of a given mass shell is
\begin{equation}
\Delta I\!=\!{4\over 3}dm_{\psi}r_{\psi}^2\!\!\left(\!{r_0\over r_{\psi}}\!\right)^4
\!\!{dr_0 \over dr_{\psi}}\!\Bigg[1\!+\!{3\over 80\pi}\!\sum_{p=1}^5Q_pB^p(p\eta_2+5)\Bigg]\!,
\label{monI8}
\end{equation}
where $\eta_2$ corresponds unambiguously to $\eta_2^{\rm (tid)}$ and
$Q_p$ is given by
\begin{equation}
Q_p={5\over p!(5-p)!}\int_0^{2\pi}\int_0^{\pi}P_2^p(\lambda)\sin^3\theta d\theta d\phi.
\label{kptid}
\end{equation}

\section{Models}\label{iscresult}

Our new version of the {\tt ATON} evolutionary code has not only many
updated and modern features, regarding the physics of stellar interior
\citep{ventura98}, but, also, is able to reproduce
stars with spherically symmetric configurations, as well as tidally and
rotationally distorted stars (see Sect.\,\ref{distortions}).
The grids cover a mass range from 0.09 to 3.8 M$_{\odot}$ and were computed
from early stages of pre-MS phase up to the main sequence. The radiative
opacities are taken from \citet{rogers1}, extended by \citet{alexander}
tables in the low-temperature regime. The OPAL equation of state 
\citep{rogers2} is used in the range 3.7$<$$\log T$$<$8.7, while in the
low-T high density regime we use the equation of state of \citet{mihalas}.
The nuclear network includes 14 elements and 22 reactions; the relevant
cross-sections are taken from \citet{caughlan}.
We adopted the solar metallicity ($Z$=0.0175, $Y$=0.27). The classical Mixing Length Theory \citep[MLT, ][]{bohm}
was used to treat the convective transport of energy, though the {\tt ATON} code
can also use the FST (Full Spectrum of Turbulence) treatment \citep{canuto96}. The mixing length parameter
has been set to $\alpha$=$\Lambda /H_p$=1.5, value 
that, according to our calibration, best reproduces the solar radius 
at the solar age with boundary conditions obtained from gray atmosphere models \citep{landin06}.

We present four sets of evolutionary models,
namely: (i) standard, spherical models (with no distorting effect) corresponding
to single non-rotating stars, (ii) binary models (distorted 
only by tidal forces) in which we treat non-rotating stars in binary systems, 
(iii) rotating models (distorted only by rotation) for representing
single rotating stars and finally (iv) rotating binary models (distorted 
simultaneously by rotation and tidal forces), useful to 
study rotating stars in binary systems.
In cases in which rotation is present, we assume rigid body rotation. 
The relation between initial angular momentum 
(J$_{\rm in}$) and stellar mass that we used was obtained 
from the respective mass-radius and
mass-moment of inertia relations from \citet{kawaler87}:
\begin{equation}
\centering
J_{\rm kaw}=1.566 \times 10^{50} \left( {M \over M_{\odot}} \right) ^{0.985}
~~~\mathrm{cgs}.
\label{kaweq}
\end{equation}
For computing the binary models (both rotating and non-rotating) we assumed
a separation of 7 times the radius of the star whose evolution is followed
(from this point on referred to as the primary), which is in the range typical for close
binary systems \citep{hilditch01}; the disturbing star is considered to be a
point mass of the same mass as its primary.
As previously stated, the pseudo-potential centrifugal terms arising from the
orbital motion around the system's center of mass are not included in the
present approximation for the tidal effects due to a companion.
We followed the evolution of internal structure constants and moment of inertia
during the pre-MS phase and tabulated them together with the corresponding
evolutionary tracks. Table\,\ref{tab100sm00} presents the 1\,M$_{\odot}$ standard model 
as an example of such tables.
Column 1 gives the logarithm of stellar age (in years); column 2, the logarithm of 
stellar luminosity (in solar units); column 3,
the logarithm of effective temperature (in K); and column 4, the logarithm of
effective gravity (in cgs). Columns 5, 6 and 7
give the logarithm of internal structure constants, and column 8 gives the gyration radius.
\begin{table}[h]
\caption{Pre-MS evolutionary tracks (including $\log k_j$ and $ \beta$) for 
1\,M$_{\odot}$ star generated with our standard models$^a$.
}
%
\vspace{0.2cm}
\label{tab100sm00}
\centering
{\scriptsize
\advance\tabcolsep by -4pt
\begin{tabular}{rrrrrrrc}
\hline \hline
${\log Age\atop{\rm (yrs)}}$ & $\log\frac{L}{L_{\odot}}$ & ${\log T_{\rm eff}\atop{\rm (K)}}$ & ${\log g\atop{\rm (cgs)}}$ & $\log k_2$ & $\log k_3$ & $\log k_4$ & $\beta$ \\ \hline
 2.2954 &    1.8909 & 3.5902 & 1.8604 & $-$0.9075 & $-$1.3965 & $-$1.7387 & 0.4177 \\ [-1.5pt]
 3.6550 &    1.6727 & 3.6057 & 2.1405 & $-$0.8654 & $-$1.3526 & $-$1.6927 & 0.4253 \\ [-1.5pt]
 4.1464 &    1.4526 & 3.6185 & 2.4119 & $-$0.8325 & $-$1.3076 & $-$1.6379 & 0.4319 \\ [-1.5pt]
 4.5435 &    1.2322 & 3.6293 & 2.6755 & $-$0.8060 & $-$1.2707 & $-$1.5920 & 0.4374 \\ [-1.5pt]
 5.1029 &    1.0251 & 3.6376 & 2.9159 & $-$0.7868 & $-$1.2434 & $-$1.5574 & 0.4415 \\ [-1.5pt]
 5.3018 &    0.8982 & 3.6418 & 3.0597 & $-$0.7776 & $-$1.2301 & $-$1.5405 & 0.4435 \\ [-1.5pt]
 5.5216 &    0.6905 & 3.6472 & 3.2889 & $-$0.7664 & $-$1.2139 & $-$1.5197 & 0.4459 \\ [-1.5pt]
 5.7953 &    0.4699 & 3.6510 & 3.5245 & $-$0.7587 & $-$1.2028 & $-$1.5053 & 0.4476 \\ [-1.5pt]
 6.0918 &    0.2493 & 3.6526 & 3.7517 & $-$0.7535 & $-$1.1954 & $-$1.4958 & 0.4488 \\ [-1.5pt]
 6.3961 &    0.0286 & 3.6516 & 3.9682 & $-$0.7769 & $-$1.2180 & $-$1.5174 & 0.4495 \\ [-1.5pt]
 6.7156 & $-$0.1866 & 3.6497 & 4.1757 & $-$0.7904 & $-$1.2420 & $-$1.5475 & 0.4458 \\ [-1.5pt]
 7.0039 & $-$0.3157 & 3.6560 & 4.3301 & $-$0.8492 & $-$1.3026 & $-$1.6126 & 0.4260 \\ [-1.5pt]
 7.2246 & $-$0.2773 & 3.6840 & 4.4038 & $-$1.0944 & $-$1.5458 & $-$1.8532 & 0.3811 \\ [-1.5pt]
 7.3696 & $-$0.1155 & 3.7255 & 4.4078 & $-$1.5590 & $-$2.0471 & $-$2.3655 & 0.3202 \\ [-1.5pt]
 7.4906 & $-$0.0389 & 3.7573 & 4.4585 & $-$1.8961 & $-$2.4599 & $-$2.8086 & 0.2812 \\ [-1.5pt]
 8.7585 & $-$0.1301 & 3.7516 & 4.5270 & $-$1.7105 & $-$2.2631 & $-$2.6077 & 0.2974 \\ [-1.5pt]
 9.3538 & $-$0.0815 & 3.7553 & 4.4933 & $-$1.8011 & $-$2.3570 & $-$2.7027 & 0.2872 \\ [-1.5pt]
 9.5876 & $-$0.0307 & 3.7587 & 4.4558 & $-$1.9017 & $-$2.4620 & $-$2.8096 & 0.2767 \\ [-1.5pt]
 9.6979 &    0.0084 & 3.7608 & 4.4252 & $-$1.9810 & $-$2.5432 & $-$2.8916 & 0.2689 \\ [-1.5pt]
 9.8058 &    0.0619 & 3.7627 & 4.3793 & $-$2.0934 & $-$2.6579 & $-$3.0078 & 0.2588 \\ \hline
\multicolumn{8}{p{0.95\columnwidth}}{$^a$The complete version of the table, including
88 tracks for the four sets of models and all the masses of Table\,\ref{zamstab00a}, will be
available only in electronic form. The tracks corresponding to rotating models
and rotating binary models will contain an additional column regarding the rotational period.}
\end{tabular}
}
\end{table}
%
%
\begin{table*}[htb]
\caption
{Internal structure constants and gyration radii for ZAMS rotating models.
See the text for details.}
\vspace{0.1cm}
\label{zamstab00a}
\centering
{\scriptsize
\advance\tabcolsep by -4.5pt
\begin{tabular}{crrrrrrcc|crrrrrrcc}
\hline \hline
\noalign{\vskip 1pt}
${{M}\atop{({\rm M}_{\odot})}}$ & log$\frac{L}{L_{\odot}}$ & ${\log T_{\rm eff}\atop{\rm (K)}}$ & ${\log g\atop{\rm (cgs)}}$ & $\log k_2$ & $\log k_3$ & $\log k_4$ & $\beta$ & ${\rm P\atop{(d)}}$ &
${{M}\atop{({\rm M}_{\odot})}}$ & log$\frac{L}{L_{\odot}}$ & $\log T_{\rm eff}$ & $\log g$ & $\log k_2$ & $\log k_3$ & $\log k_4$ & $\beta$ & ${\rm P\atop{(d)}}$ \\ [4pt]
\hline
\hline
\multicolumn{8}{c}{standard (spherically symmetric) models} &&
\multicolumn{9}{c}{tidally (non-rotating) distorted models}\\
\hline
0.09 & $-$3.3197 &  3.4367 &  5.4113 & $-$0.8134 & $-$1.2455 & $-$1.5272 &  0.4600 & --- &
0.09 & $-$3.3260 &  3.4355 &  5.4146 & $-$0.8173 & $-$1.2500 & $-$1.5318 &  0.4601 & --- \\ [-1.7pt]
0.10 & $-$3.0098 &  3.4839 &  5.3360 & $-$0.8359 & $-$1.2743 & $-$1.5613 &  0.4554 & --- &
0.10 & $-$3.0131 &  3.4836 &  5.3398 & $-$0.8391 & $-$1.2779 & $-$1.5651 &  0.4556 & --- \\ [-1.7pt]
0.20 & $-$2.2529 &  3.5255 &  5.0462 & $-$0.8943 & $-$1.3520 & $-$1.6567 &  0.4453 & --- &
0.20 & $-$2.2540 &  3.5255 &  5.0490 & $-$0.9000 & $-$1.3592 & $-$1.6651 &  0.4454 & --- \\ [-1.7pt]
0.30 & $-$1.9334 &  3.5444 &  4.9788 & $-$0.8720 & $-$1.3253 & $-$1.6269 &  0.4489 & --- &
0.30 & $-$1.9344 &  3.5445 &  4.9815 & $-$0.8770 & $-$1.3317 & $-$1.6345 &  0.4490 & --- \\ [-1.7pt]
0.40 & $-$1.6724 &  3.5626 &  4.9153 & $-$0.9176 & $-$1.3812 & $-$1.6922 &  0.4402 & --- &
0.40 & $-$1.6716 &  3.5628 &  4.9168 & $-$0.9213 & $-$1.3870 & $-$1.6995 &  0.4397 & --- \\ [-1.7pt]
0.50 & $-$1.3820 &  3.5885 &  4.8254 & $-$1.0460 & $-$1.5023 & $-$1.8099 &  0.4119 & --- &
0.50 & $-$1.3815 &  3.5888 &  4.8275 & $-$1.0501 & $-$1.5082 & $-$1.8173 &  0.4114 & --- \\ [-1.7pt]
0.60 & $-$1.0632 &  3.6252 &  4.7325 & $-$1.2494 & $-$1.7065 & $-$2.0127 &  0.3808 & --- &
0.60 & $-$1.0598 &  3.6240 &  4.7260 & $-$1.2309 & $-$1.6897 & $-$1.9975 &  0.3821 & --- \\ [-1.7pt]
0.70 & $-$0.7584 &  3.6639 &  4.6496 & $-$1.4603 & $-$1.9300 & $-$2.2404 &  0.3534 & --- &
0.70 & $-$0.7558 &  3.6640 &  4.6487 & $-$1.4621 & $-$1.9334 & $-$2.2449 &  0.3532 & --- \\ [-1.7pt]
0.80 & $-$0.4808 &  3.7026 &  4.5849 & $-$1.6441 & $-$2.1288 & $-$2.4442 &  0.3314 & --- &
0.80 & $-$0.4778 &  3.7024 &  4.5825 & $-$1.6417 & $-$2.1282 & $-$2.4453 &  0.3312 & --- \\ [-1.7pt]
0.90 & $-$0.3454 &  3.7280 &  4.6022 & $-$1.7067 & $-$2.2259 & $-$2.5554 &  0.3216 & --- &
0.90 & $-$0.2348 &  3.7312 &  4.5059 & $-$1.8397 & $-$2.3484 & $-$2.6735 &  0.3071 & --- \\ [-1.7pt]
1.00 & $-$0.1253 &  3.7540 &  4.5319 & $-$1.9037 & $-$2.4571 & $-$2.8011 &  0.2957 & --- &
1.00 & $-$0.1228 &  3.7546 &  4.5331 & $-$1.9185 & $-$2.4769 & $-$2.8234 &  0.2950 & --- \\ [-1.7pt]
1.20 &    0.3184 &  3.7935 &  4.3251 & $-$2.4573 & $-$3.1435 & $-$3.5662 &  0.2361 & --- &
1.20 &    0.3319 &  3.7941 &  4.3156 & $-$2.4968 & $-$3.1894 & $-$3.6146 &  0.2334 & --- \\ [-1.7pt]
1.40 &    0.6830 &  3.8214 &  4.1394 & $-$2.8859 & $-$3.7014 & $-$4.2406 &  0.1962 & --- &
1.40 &    0.6901 &  3.8219 &  4.1355 & $-$2.9077 & $-$3.7278 & $-$4.2709 &  0.1950 & --- \\ [-1.7pt]
1.60 &    0.9558 &  3.8590 &  4.0749 & $-$2.9992 & $-$3.8561 & $-$4.4404 &  0.1861 & --- &
1.60 &    0.9630 &  3.8595 &  4.0709 & $-$3.0239 & $-$3.8827 & $-$4.4670 &  0.1853 & --- \\ [-1.7pt]
1.80 &    1.1839 &  3.8998 &  4.0611 & $-$2.9891 & $-$3.8436 & $-$4.4275 &  0.1869 & --- &
1.80 &    1.1883 &  3.9011 &  4.0629 & $-$3.0035 & $-$3.8618 & $-$4.4467 &  0.1866 & --- \\ [-1.7pt]
2.00 &    1.3797 &  3.9357 &  4.0547 & $-$2.9667 & $-$3.8203 & $-$4.4044 &  0.1887 & --- &
2.00 &    1.3851 &  3.9366 &  4.0541 & $-$2.9796 & $-$3.8339 & $-$4.4182 &  0.1882 & --- \\ [-1.7pt]
2.30 &    1.6345 &  3.9811 &  4.0419 & $-$2.9337 & $-$3.7823 & $-$4.3644 &  0.1913 & --- &
2.30 &    1.6376 &  3.9826 &  4.0464 & $-$2.9370 & $-$3.7882 & $-$4.3727 &  0.1912 & --- \\ [-1.7pt]
2.50 &    1.7839 &  4.0073 &  4.0339 & $-$2.9098 & $-$3.7554 & $-$4.3376 &  0.1930 & --- &
2.50 &    1.7867 &  4.0089 &  4.0385 & $-$2.9233 & $-$3.7718 & $-$4.3542 &  0.1930 & --- \\ [-1.7pt]
2.80 &    1.9782 &  4.0438 &  4.0349 & $-$2.8593 & $-$3.7029 & $-$4.2840 &  0.1967 & --- &
2.80 &    1.9834 &  4.0447 &  4.0343 & $-$2.8732 & $-$3.7160 & $-$4.2976 &  0.1962 & --- \\ [-1.7pt]
3.00 &    2.0976 &  4.0649 &  4.0296 & $-$2.8407 & $-$3.6809 & $-$4.2612 &  0.1985 & --- &
3.00 &    2.1002 &  4.0664 &  4.0344 & $-$2.8480 & $-$3.6905 & $-$4.2709 &  0.1985 & --- \\ [-1.7pt]
3.30 &    2.2587 &  4.0938 &  4.0257 & $-$2.8049 & $-$3.6406 & $-$4.2186 &  0.2014 & --- &
3.30 &    2.2589 &  4.0959 &  4.0350 & $-$2.7987 & $-$3.6360 & $-$4.2145 &  0.2018 & --- \\ [-1.7pt]
3.50 &    2.3542 &  4.1121 &  4.0287 & $-$2.7741 & $-$3.6065 & $-$4.1826 &  0.2037 & --- &
3.50 &    2.3593 &  4.1128 &  4.0275 & $-$2.7974 & $-$3.6341 & $-$4.2126 &  0.2032 & --- \\ [-1.7pt]
3.80 &    2.4869 &  4.1368 &  4.0307 & $-$2.7394 & $-$3.5731 & $-$4.1496 &  0.2069 & --- &
3.80 &    2.4915 &  4.1378 &  4.0313 & $-$2.7537 & $-$3.5871 & $-$4.1636 &  0.2065 & --- \\ \hline
\hline
\multicolumn{8}{c}{rotationally distorted (single star) models} &&
\multicolumn{9}{c}{rotationally and tidally distorted models} \\
\hline
0.09 & $-$3.5403 & 3.3830 & 5.4499 & $-$0.8321 & $-$1.2660 & $-$1.5486 & 0.4657 & 0.0323 & 
0.09 & $-$3.5459 & 3.3818 & 5.4520 & $-$0.8317 & $-$1.2657 & $-$1.5481 & 0.4655 & 0.0322 \\ [-1.7pt]
0.10 & $-$3.2999 & 3.4264 & 5.4644 & $-$0.8653 & $-$1.3053 & $-$1.5923 & 0.4666 & 0.0237 &
0.10 & $-$3.3060 & 3.4251 & 5.4666 & $-$0.8649 & $-$1.3047 & $-$1.5916 & 0.4664 & 0.0236 \\ [-1.7pt]
0.20 & $-$2.3178 & 3.5083 & 5.1457 & $-$0.9989 & $-$1.4770 & $-$1.7971 & 0.4450 & 0.0430 &
0.20 & $-$2.3170 & 3.5086 & 5.1456 & $-$0.9965 & $-$1.4741 & $-$1.7938 & 0.4449 & 0.0429 \\ [-1.7pt]
0.30 & $-$1.9701 & 3.5350 & 5.0252 & $-$0.9202 & $-$1.3823 & $-$1.6909 & 0.4520 & 0.0777 &
0.30 & $-$1.9707 & 3.5349 & 5.0269 & $-$0.9194 & $-$1.3815 & $-$1.6901 & 0.4518 & 0.0775 \\ [-1.7pt]
0.40 & $-$1.7068 & 3.5551 & 4.9499 & $-$0.9340 & $-$1.4047 & $-$1.7203 & 0.4465 & 0.1159 &
0.40 & $-$1.7041 & 3.5553 & 4.9491 & $-$0.9358 & $-$1.4072 & $-$1.7232 & 0.4457 & 0.1160 \\ [-1.7pt]
0.50 & $-$1.4194 & 3.5808 & 4.8578 & $-$1.0457 & $-$1.5085 & $-$1.8213 & 0.4182 & 0.1561 &
0.50 & $-$1.4171 & 3.5810 & 4.8577 & $-$1.0438 & $-$1.5063 & $-$1.8189 & 0.4176 & 0.1562 \\ [-1.7pt]
0.60 & $-$1.1005 & 3.6164 & 4.7590 & $-$1.2446 & $-$1.7069 & $-$2.0180 & 0.3852 & 0.1995 &
0.60 & $-$1.0981 & 3.6165 & 4.7583 & $-$1.2424 & $-$1.7044 & $-$2.0152 & 0.3849 & 0.2001 \\ [-1.7pt]
0.70 & $-$0.7913 & 3.6551 & 4.6710 & $-$1.4587 & $-$1.9288 & $-$2.2416 & 0.3566 & 0.2445 &
0.70 & $-$0.7897 & 3.6550 & 4.6706 & $-$1.4540 & $-$1.9246 & $-$2.2378 & 0.3565 & 0.2454 \\ [-1.7pt]
0.80 & $-$0.5102 & 3.6945 & 4.6054 & $-$1.6515 & $-$2.1405 & $-$2.4603 & 0.3334 & 0.2845 &
0.80 & $-$0.5075 & 3.6935 & 4.6000 & $-$1.6403 & $-$2.1236 & $-$2.4405 & 0.3339 & 0.2897 \\ [-1.7pt]
0.90 & $-$0.2657 & 3.7247 & 4.5344 & $-$1.8640 & $-$2.3763 & $-$2.7039 & 0.3084 & 0.3243 &
0.90 & $-$0.2638 & 3.7248 & 4.5342 & $-$1.8639 & $-$2.3754 & $-$2.7025 & 0.3080 & 0.3247 \\ [-1.7pt]
1.00 & $-$0.1458 & 3.7465 & 4.5486 & $-$1.9242 & $-$2.4777 & $-$2.8225 & 0.2963 & 0.3233 &
1.00 & $-$0.1444 & 3.7467 & 4.5495 & $-$1.9231 & $-$2.4757 & $-$2.8197 & 0.2959 & 0.3228 \\ [-1.7pt]
1.20 &    0.2881 & 3.7838 & 4.3569 & $-$2.4765 & $-$3.1529 & $-$3.5633 & 0.2362 & 0.3968 &
1.20 &    0.3355 & 3.7672 & 4.2349 & $-$2.4134 & $-$3.0060 & $-$3.3663 & 0.2432 & 0.5464 \\ [-1.7pt]
1.40 &    0.6679 & 3.8070 & 4.1585 & $-$2.9872 & $-$3.8173 & $-$4.3553 & 0.1891 & 0.4931 &
1.40 &    0.6753 & 3.8065 & 4.1514 & $-$3.0084 & $-$3.8409 & $-$4.3800 & 0.1878 & 0.4974 \\ [-1.7pt]
1.60 &    0.9437 & 3.8406 & 4.0753 & $-$3.1595 & $-$4.0570 & $-$4.6681 & 0.1749 & 0.5857 &
1.60 &    0.9492 & 3.8400 & 4.0696 & $-$3.1646 & $-$4.0617 & $-$4.6729 & 0.1740 & 0.5902 \\ [-1.7pt]
1.80 &    1.1741 & 3.8852 & 4.0602 & $-$3.1158 & $-$4.0057 & $-$4.6137 & 0.1778 & 0.6840 &
1.80 &    1.1772 & 3.8852 & 4.0587 & $-$3.1232 & $-$4.0129 & $-$4.6212 & 0.1774 & 0.6854 \\ [-1.7pt]
2.00 &    1.3757 & 3.9229 & 4.0452 & $-$3.0779 & $-$3.9557 & $-$4.5578 & 0.1807 & 0.7948 &
2.00 &    1.3793 & 3.9225 & 4.0415 & $-$3.0858 & $-$3.9640 & $-$4.5662 & 0.1801 & 0.7992 \\ [-1.7pt]
2.30 &    1.6297 & 3.9723 & 4.0388 & $-$3.0079 & $-$3.8755 & $-$4.4723 & 0.1858 & 0.9586 &
2.30 &    1.6333 & 3.9719 & 4.0353 & $-$3.0205 & $-$3.8886 & $-$4.4851 & 0.1852 & 0.9638 \\ [-1.7pt]
2.50 &    1.7790 & 4.0003 & 4.0331 & $-$2.9768 & $-$3.8396 & $-$4.4328 & 0.1886 & 1.0773 &
2.50 &    1.7806 & 4.0007 & 4.0342 & $-$2.9814 & $-$3.8443 & $-$4.4376 & 0.1884 & 1.0762 \\ [-1.7pt]
2.80 &    1.9768 & 4.0378 & 4.0290 & $-$2.9217 & $-$3.7756 & $-$4.3644 & 0.1927 & 1.2590 &
2.80 &    1.9787 & 4.0380 & 4.0296 & $-$2.9249 & $-$3.7803 & $-$4.3700 & 0.1925 & 1.2588 \\ [-1.7pt]
3.00 &    2.0944 & 4.0602 & 4.0286 & $-$2.8859 & $-$3.7392 & $-$4.3279 & 0.1954 & 1.3821 &
3.00 &    2.0957 & 4.0607 & 4.0309 & $-$2.8819 & $-$3.7328 & $-$4.3199 & 0.1954 & 1.3785 \\ [-1.7pt]
3.30 &    2.2542 & 4.0906 & 4.0287 & $-$2.8342 & $-$3.6800 & $-$4.2646 & 0.1993 & 1.5722 &
3.30 &    2.2577 & 4.0904 & 4.0260 & $-$2.8433 & $-$3.6885 & $-$4.2723 & 0.1988 & 1.5784 \\ [-1.7pt]
3.50 &    2.3530 & 4.1085 & 4.0259 & $-$2.8139 & $-$3.6572 & $-$4.2405 & 0.2014 & 1.7091 &
3.50 &    2.3527 & 4.1094 & 4.0311 & $-$2.8084 & $-$3.6511 & $-$4.2333 & 0.2016 & 1.6983 \\ [-1.7pt]
3.80 &    2.4866 & 4.1338 & 4.0271 & $-$2.7748 & $-$3.6134 & $-$4.1936 & 0.2048 & 1.9082 &
3.80 &    2.4880 & 4.1343 & 4.0291 & $-$2.7711 & $-$3.6083 & $-$4.1878 & 0.2047 & 1.9038 \\ \hline
\end{tabular}
}
\end{table*}

Table~\ref{zamstab00a} presents the internal structure constants
and the gyration radius at the ZAMS, for both non-rotating and rotating models.
Its successive columns give the stellar mass (in $M_{\odot}$);
the logarithm of the stellar luminosity (in solar units); the logarithm 
of the effective temperature (in K); the logarithm of the surface
gravity (in cgs units); the logarithm of the
internal structure constants $k_2$, $k_3$ and $k_4$;
the radius of gyration $\beta$ (cf. Eq.\,\ref{beta});
and, finally, the rotational periods (in days), 
when rotation is present.

\section{Results and discussion}\label{iscdisc}

In this section, we discuss the results yielded by the models presented in
Sect.\,(\ref{iscresult}) and the differences obtained with each set of models.
Then we proceed with a comparison with other works in the literature.

\subsection{Effects on the internal structure constants}

As previously mentioned, \citet{mohan90} found that the effects of
rotational distortions on main-sequence stellar models are greater than those of tidal
distortions, though their results were obtained only for binary models with a mass
ratio of 0.1. We investigated the ZAMS models in order
to verify such differences, by calculating how different the
non-standard values of the second-order internal structure constants are as compared
with the standard ones for each stellar mass at the ZAMS. The binary models produced
values of $\log k_2$ lower than the standard models, on average, by a difference of about 0.0170, with a maximum
difference of 0.1330 occurring for the 0.9\,M$_{\odot}$ model. For the rotating
and the rotating binary models this average difference is 0.0570 
and 0.0594, respectively; the maximum difference happens
at the mass of  1.6\,M$_{\odot}$ for both sets of models, being 0.1603 for
rotating models and 0.1654 for rotating binary ones.

To better assess the relative importance of the rotational and tidal
effects, we also ran rotating binary models of 0.3 and 1.0 solar masses with (a) separations
of 3, 14, 35 and 70 times the radius of the primary, (b) masses of the
disturbing star corresponding to the double and half values of the primary 
mass, and (c) initial angular momentum equal to 0, 1
and 3 J$_{\rm kaw}$, where J$_{\rm kaw}$ is given in 
Eq.~(\ref{kaweq}). Table\,\ref{newcomp} presents, from left to right, the primary mass, the
initial angular momentum, the secondary mass, the separation between the components and the
logarithm of $k_2$ at the ZAMS
for these additional models. From that table we can see that, with the only exception of
the case of 1\,M$_{\odot}$ with orbital separation of 3 times the primary's radius and
$J_{\rm in}$ equal to J$_{\rm kaw}$,
$\log k_2$ is mainly affected by the inclusion of rotational effects. For 0.3\,M$_{\odot}$, considering
wider orbits and assuming different mass binary components do not alter significantly
the $\log k_2$ values. For the 1\,M$_{\odot}$ model, $\log k_2$ decreases slightly when
the distance between the stars increases, except for separations 
of 3 and 70 stellar radius, and when we vary the mass of the
companion stars. The maximum difference occurs for 1\,M$_{\odot}$ models, when we
consider a companion star of 2\,M$_{\odot}$.

\begin{table}[!h]
\caption{Values of $\log k_2$ obtained for additional models$^b$.}
\vspace{0.4cm}
\label{newcomp}  
\vspace{-0.2cm}
\centering
\begin{tabular}{ccccc}
\hline \hline
M$_{\rm prim}$ (M$_{\odot}$) & J$_{\rm in}$ (J$_{\rm kaw}$) & M$_{\rm sec}$ (M$_{\odot}$) & R (r$_{\star}$) & $\log$(k$_2$) \\ \hline      
0.30 & \dots     & \dots    & \dots    & $-$0.872 \\[-1pt]
0.30 & \dots     & 0.30     & 7        & $-$0.877 \\[-1pt]
0.30 & 1         & \dots    & \dots    & $-$0.921 \\[-1pt]
0.30 & 1         & 0.30     & 7        & $-$0.919 \\[-1pt]
0.30 & 1         & 0.30     & 14       & $-$0.918 \\[-1pt]
0.30 & 1         & 0.30     & 35       & $-$0.919 \\[-1pt]
0.30 & 1         & 0.30     & 70       & $-$0.920 \\[-1pt]
0.30 & 1         & 0.15     & 7        & $-$0.918 \\[-1pt]
0.30 & 1         & 0.60     & 7        & $-$0.919 \\[-1pt]
0.30 & \dots     & 0.30     & 3        & $-$0.868 \\[-1pt]
0.30 & 1         & 0.30     & 3        & $-$0.916 \\[-1pt]
1.00 & \dots     & \dots    & \dots    & $-$1.904 \\[-1pt]
1.00 & \dots     & 1.00     & 7        & $-$1.918 \\[-1pt]
1.00 & 1         & \dots    & \dots    & $-$1.924 \\[-1pt]
1.00 & 1         & 1.00     & 7        & $-$1.923 \\[-1pt]
1.00 & 1         & 1.00     & 14       & $-$1.926 \\[-1pt]
1.00 & 1         & 1.00     & 35       & $-$1.935 \\[-1pt]
1.00 & 1         & 1.00     & 70       & $-$1.930 \\[-1pt]
1.00 & 1         & 0.50     & 7        & $-$1.932 \\[-1pt]
1.00 & 1         & 2.00     & 7        & $-$1.939 \\[-1pt]
1.00 & \dots     & 1.00     & 3        & $-$2.094 \\[-1pt]
1.00 & 1         & 1.00     & 3        & $-$2.027 \\[-1pt]
1.00 & 3         & 1.00     & 3        & $-$2.277 \\ \hline
\multicolumn{5}{p{0.95\columnwidth}}{$^b$Ellipsis dots (\dots) 
indicate missing values, because the model is either non-rotating or not 
binary. Rows with all numbers correspond to rotating binary models.}
\end{tabular}
\end{table}

\begin{figure*}[t]
\includegraphics[width=\textwidth]{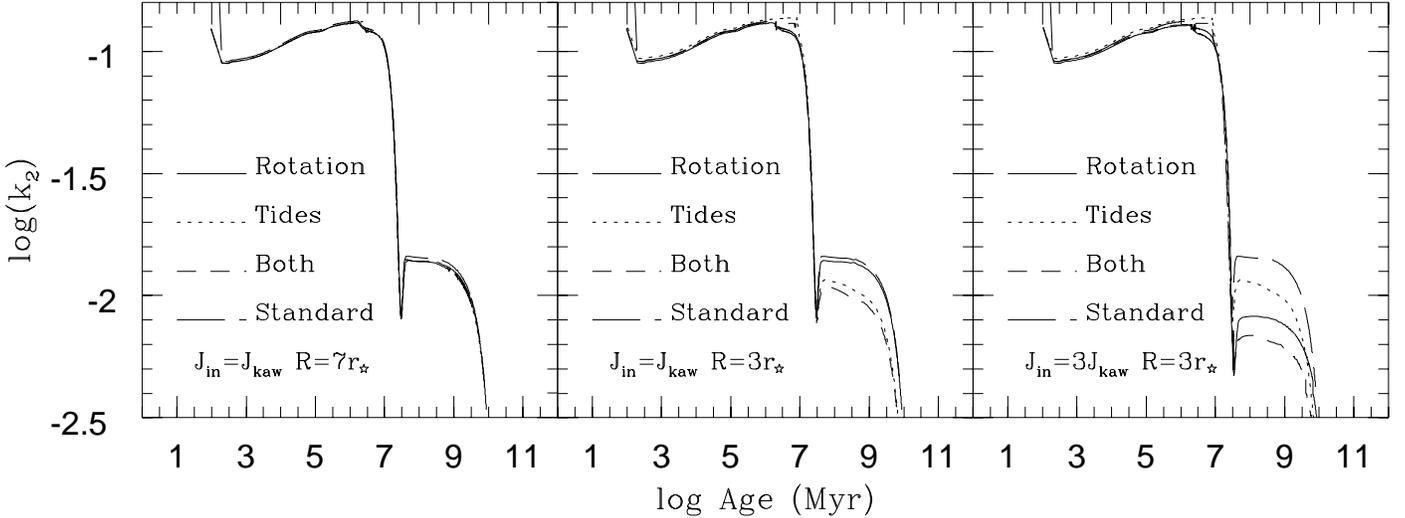}
\caption{$\log k_2$ values against age for an 1\,M$_{\odot}$ models
with orbital separation $R$ and initial angular momentum $J_{\rm in}$ as indicated, where
$r_\star$ denotes the primary's radius and $J_{\rm kaw}$ the initial angular momentum given by
Kawaler's \citeyearpar{kawaler87} relations. Long-dashed lines represent values for the
standard (non-rotating, single star) model; solid lines, the rotating-only model; dotted lines,
the binary model without rotation; and the short-dashed lines, the rotating binary model.}
\label{fig3R}
\end{figure*}

The particular cases of 1\,M$_{\odot}$ with orbital separation of 3
times the primary's radius and $J_{\rm in}\!=\!J_{\rm kaw}$ or \mbox{$J_{\rm in}\!=\!3J_{\rm kaw}$}
(where $J_{\rm kaw}$ denotes the initial angular momentum given by Eq.\,\ref{kaweq}) demonstrate
however that, depending on the star's rotation rate, a shorter orbital
separation can make the tidal effects to overcome the rotational ones regarding the
internal structure constants. This can be seen in Fig.\,(\ref{fig3R}), which also shows
that the differences become more apparent for ages starting at the ZAMS.
%

For the gyration radii, we found, for a given mass at the ZAMS, values slightly higher for
the rotating binary than for those calculated with the standard models. This is due to the
combined differences between the radius and the rotational inertia of each model.

Concerning the effects on the internal structure of the stars, the
models distorted only by tidal forces are in general similar to the standard ones and
the models distorted only by rotation are similar to the rotating binary
models. This is due to the fact that, for the adopted orbital separation, the rotation effects
are, on average, more important than the tidal ones.
Here, we concentrate in the differences between the standard models and the rotating binary
models, since this is the case of scientific interest for apsidal motion studies.

\begin{figure}
\includegraphics[width=\columnwidth]{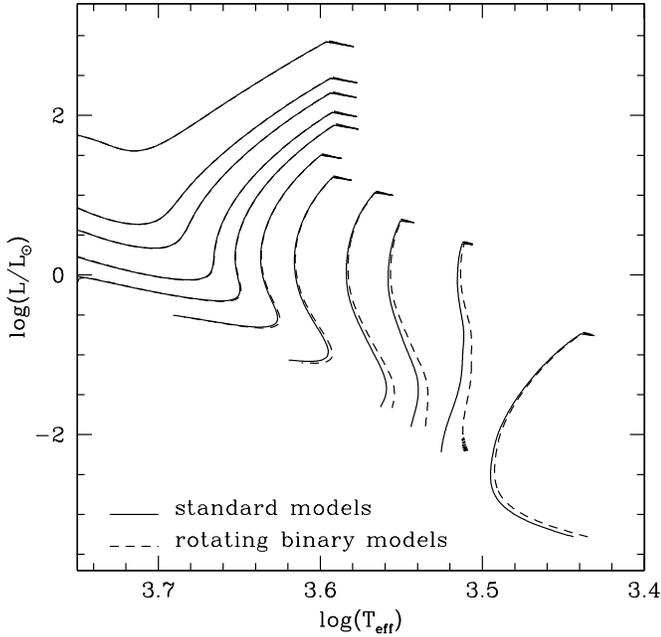}
\caption{Evolutionary tracks for standard and for rotating binary models.
The model masses correspond to 0.09, 0.2 0.3, 0.4, 0.6, 0.8,
1.0, 1.2, 1.8, 2.0, and 3.8\,M$_{\odot}$ from bottom to top.}
\label{evol0011left}
\end{figure}

In Fig.\,\ref{evol0011left} we show the path followed by our standard models (solid lines)
and our rotationally and tidally distorted models (dashed lines)
for the masses of 0.09, 0.2, 0.3, 0.4, 0.6, 0.8, 1.0, 1.2, 1.8, 2.0, and 3.8\,M$_{\odot}$.
We can see that the distorted pre-MS evolutionary tracks have lower
$T_{\rm eff}$ than their standard counterparts, as it is already known
for stars distorted only by rotation. 
\citet{sack70} reported that
the distorting effects of rotation affects late-type, main-sequence stars
much more seriously than the early-type ones,
and that this behavior could be approximately 
understood by looking at the variation of the power law expressions 
of the nuclear-energy-generation rates along the main sequence.
This trend is also observed when tidal and rotational distortions
act together during the pre-MS.
%

\begin{figure}[t]
\centering{
\includegraphics[width=\columnwidth]{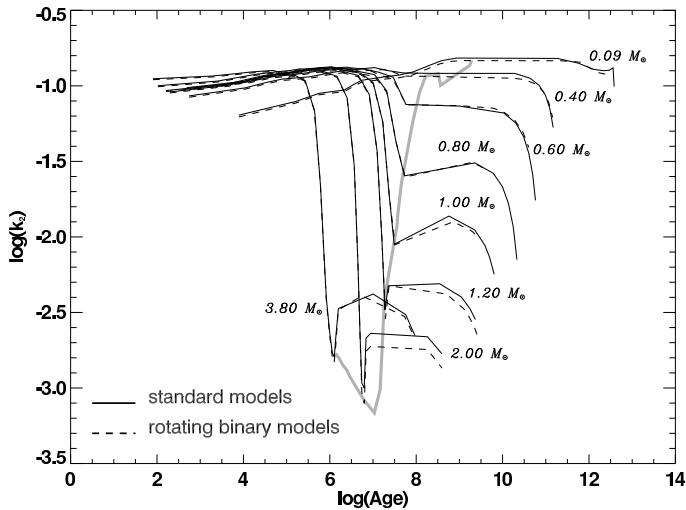}}
\caption{The time evolution of $\log k_2$ for selected
masses of the standard and the rotating binary models. 
The gray line corresponds to the ZAMS.
}
\label{kjagem0011}
\end{figure}

Fig.\,\ref{kjagem0011}, in which we plot $\log k_2$ against the
stellar age for selected masses of our standard models and rotating binary models,
shows the significant change in mass concentration during
the pre-MS evolution, especially for higher masses. 
For ages less than 1\,Myr the $\log k_2$ do not vary significantly, neither
with time nor with mass (see Fig.\,\ref{kjagem0011}). For models with masses
lower than 0.3\,M$_{\odot}$, that correspond to the mass
range where stars are almost completely convective, $\log k_2$
remains roughly constant during the evolution. For masses greater than
0.3\,M$_{\odot}$ the values of $\log k_2$ are constant until a given age,
after which they start to drop; the decrease of $\log k_2$ starts earlier
as the mass increases.  It seems that $\log k_2$ remains constant until
the star develops a radiative core.
%
%
\begin{figure}[t]
\centering{
\includegraphics[width=\columnwidth]{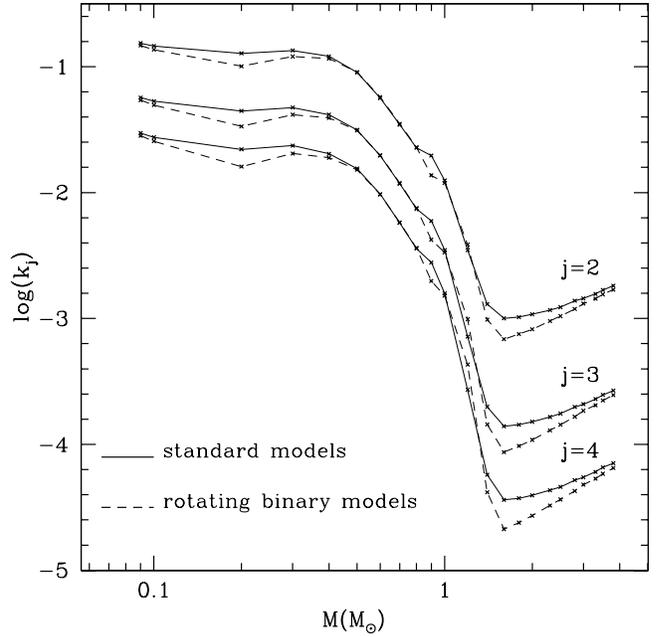}}
\caption{$\log k_j$ as a function of $\log M$ for ZAMS models, for the
the same set of masses as in Fig.\, \ref{evol0011left}.
}
\label{kjmassm0011}
\end{figure}

The values of $\log k_j$ versus stellar mass for standard models and rotating binary models
at the ZAMS are shown in Fig.\,\ref{kjmassm0011}, in which we see that
the distorted models corresponding curves remain
below those corresponding to the standard models. 
The values of $\log k_j$ are more sensitive 
to the distorting effects for $M$$<$0.5\,M$_{\odot}$
and for $M$$>$1.4\,M$_{\odot}$. The values of $\log k_2$, $\log k_3$, and
$\log k_4$ remain roughly constant in the mass range 0.09-0.4\,M$_{\odot}$. 
As the mass increases from 0.4\,M$_{\odot}$ to 1.5\,M$_{\odot}$, $k_j$ drop
significantly (2-3 orders of magnitude) and reach their minimum value at
1.5\,M$_{\odot}$. For masses greater than 1.5\,M$_{\odot}$ we note a
parallel behavior of the $k_j$. 
H87 pointed out that, for a given mass, there is a significant
decrease in $k_j$ for increasing $j$, and that the relevance of including the
higher order terms, when comparing with observations, can be judged from
Eq.\,(\ref{isc234}). In our analysis $\log k_3$ and $\log k_4$ are really less important
than $\log k_2$ in the mass range that he analyzed (0.5-32\,M$_{\odot}$),
but the same statement cannot be extended to less massive stars
(M$\leq$0.5\,M$_{\odot}$): as can be seen from Table \ref{zamstab00a} and also from Fig.\,\ref{kjmassm0011},
$k_4$ is lower than $k_2$ by more than one order of magnitude in the mass range 1.5-3.8\,M$_{\odot}$, but
this difference decreases to less than one order of magnitude for stars less massive
than  0.5\,M$_{\odot}$. So, in the low-mass range, the assumption that the
harmonics of order greater than $j$=2 can be neglected, widely used when
analyzing the apsidal motion of binary systems, seems not to be entirely
justified, except maybe when computing the apsidal motion rate
$\dot{\omega}$ in which, besides a dependence on $\log k_j$, 
there is also a dependence on $(r_0/R)^{j+3}$ that will
make the $j\!=\!2$ term always dominate unless $r_0$ is very close to $R$.


\begin{figure}[t]
\centering{
\includegraphics[width=\columnwidth]{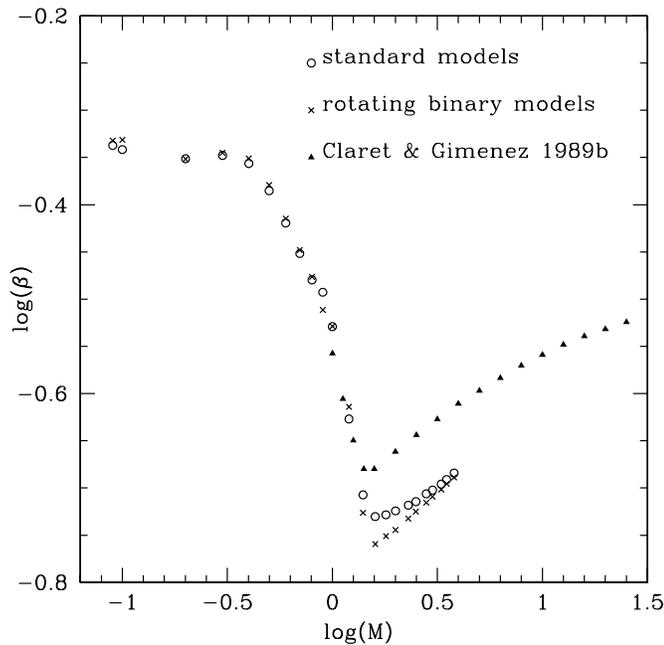}}
\caption {Standard models (open circles) and rotating binary models (crosses)
in the $\log \beta$ - $\log M$ plane. The full triangles 
correspond to the results by \citet{claret89b}.
}
\label{btmk20011}
\end{figure}
 
In Fig.\,\ref{btmk20011} we show $\log \beta$ as a function of $\log M$, at
the ZAMS, for our standard and non-standard models and also for the non-rotating 
models by \citet{claret89b}. For masses above 1.4\,M$_{\odot}$ our models predict a
lower value of $\log \beta$ than the \citet{claret89b} ones; the values
predicted by our rotating binary models are even lower.
Although these models have slightly different
chemical compositions, the comparison is still valid. A control made with a one solar mass model with
the same initial chemical composition as those by \citet{claret89b} shows that
the gyration radius at the ZAMS did not vary significantly.

The time evolution of $\log \beta$ has the same behavior as the time
evolution of $\log k_j$ (Fig.\,\ref{kjagem0011}).  Also, $\log \beta$ behaves
roughly in the same way as $\log k_j$ as a function of $\log M$ (Fig.\,\ref{kjmassm0011}),
with a minimum at 1.5\,M$_{\odot}$, corresponding to the change in the dominant
energy source from the p-p chain to the CNO cycle \citep{claret89b}. This
similarity is due to the known linear relationship between $\log k_j$ and
$\log \beta$, also seen in Fig.\,1 by \citet{motz52}.

\begin{figure*}[!t]
\centering{
\includegraphics[width=5.7cm]{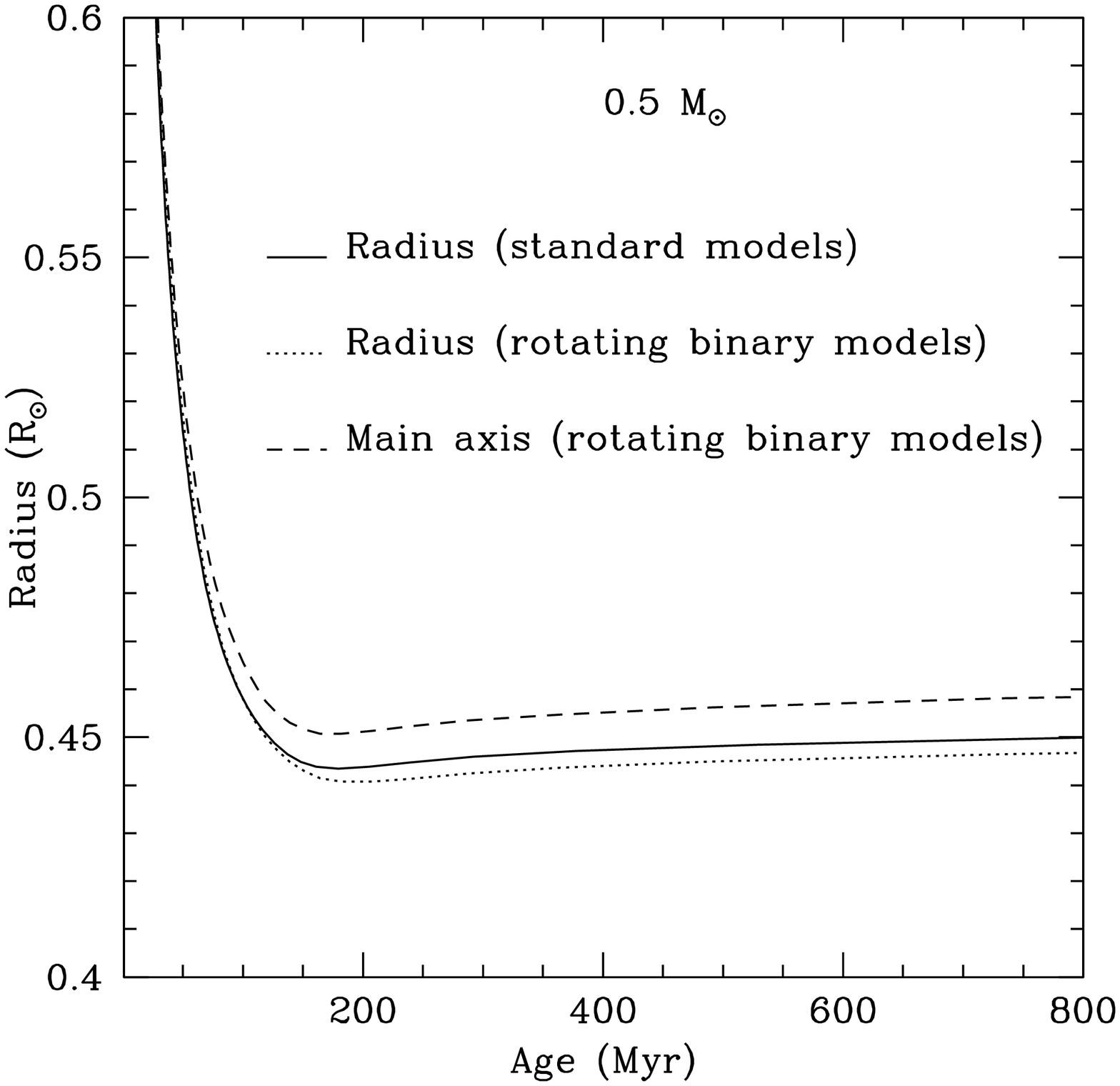}
\includegraphics[width=5.7cm]{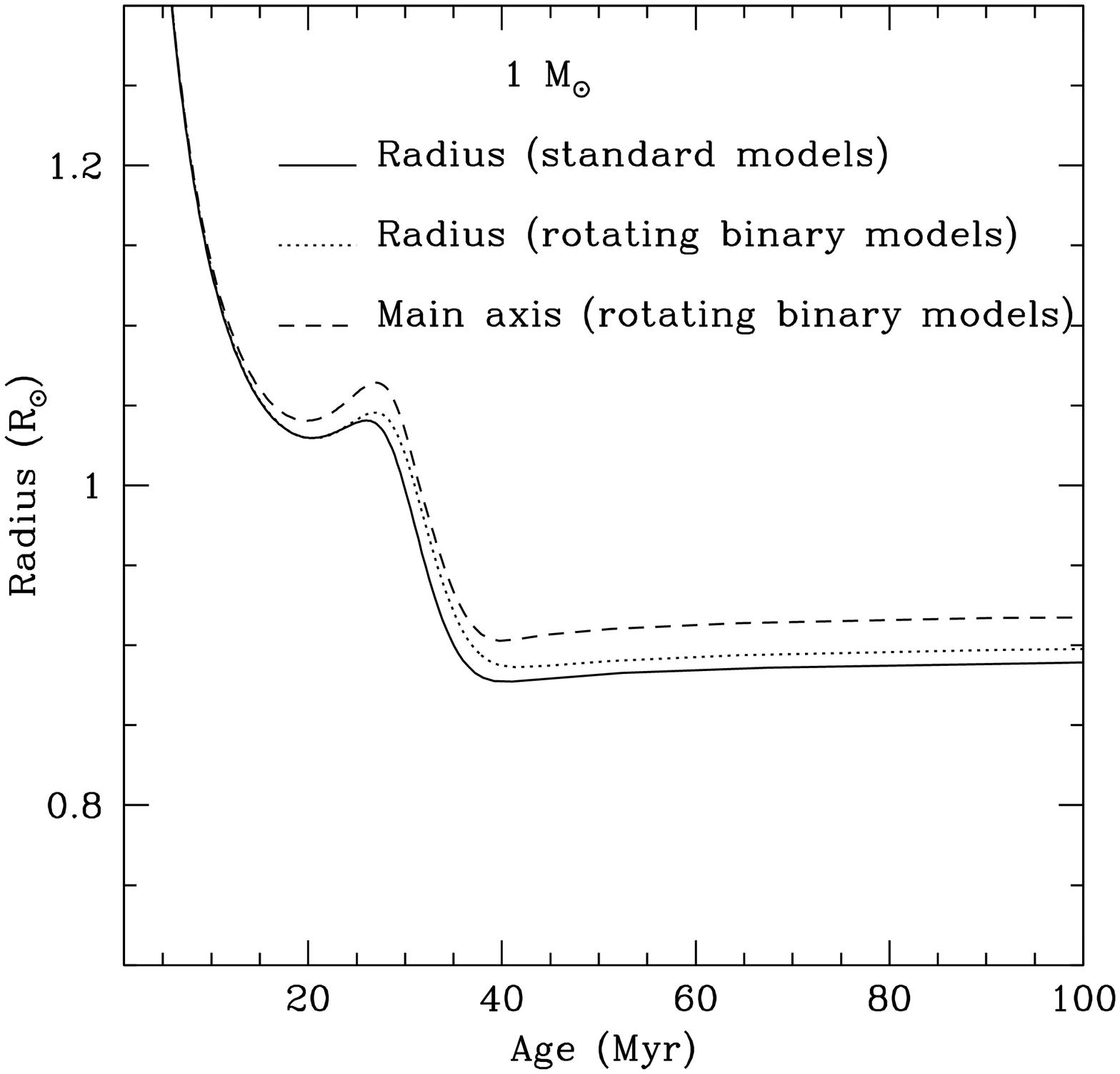}
\includegraphics[width=5.7cm]{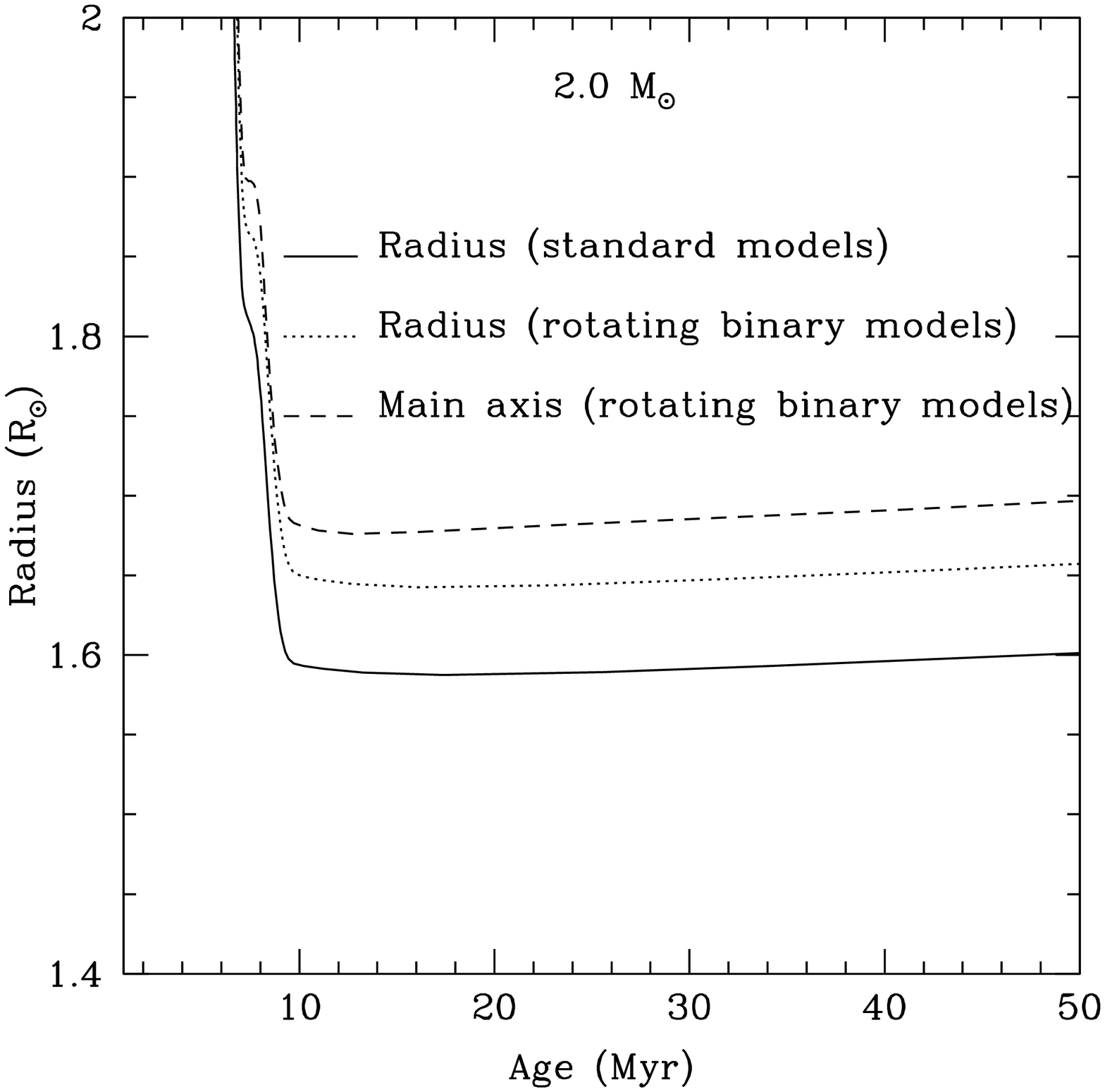}}
\caption{Stellar radius and main axis of our standard models and rotating binary models.
We report only the results for 0.5,
1.0 and 2.0 M$_{\odot}$ models. The solid and dotted lines denote the radius of standard 
models and rotating binary models, respectively, while the dashed line 
stand for the radius along the main axis of the rotating binary models.}
\label{radcomp}
\end{figure*}
The differences of the stellar radii as a function of stellar age and mass
for the standard and distorted models are shown in  Fig.\,\ref{radcomp}.
For a 0.5\,M$_{\odot}$ star, it can be seen that the stellar radii at the
ZAMS of the rotating binary models are slightly smaller than those produced by the standard models.
For the 1\,M$_{\odot}$ model, the situation is the opposite: the
distorted models predict a slightly larger stellar radius at 
the ZAMS than the standard models; this difference becomes
still larger for the 2\,M$_{\odot}$ model.
The threshold mass for this transition is at {$\sim$}%
0.7\,M$_{\odot}$.
\citet{sack70} had already noted this behavior,
finding a threshold mass about 1.3-1.5\,M$_{\odot}$ and suggested that it
could be related to the cross-over from the proton-proton
chain to the CNO cycle that occurs around 1.5\,M$_{\odot}$, depending
on the initial chemical composition.
According to him, this behavior of the effects of rotation 
shows up in all physical quantities of a star. 
Fig.\,\ref{radcomp} also shows the stellar radius along the mean system
axis, defined as the axis joining the two stars. For all models computed
the mean axis is greater than the mean stellar radius during the whole evolution,
the difference being more evident for ages higher than $\sim$100\,Myr (0.5\,M$_{\odot}$),
$\sim$20\,Myr (1.0\,M$_{\odot}$) and $\sim$10\,Myr (2.0\,M$_{\odot}$).  

As a final remark, we verified that tidal effects act in the same way
as rotational ones but on a smaller scale.

\subsection{Comparison with other works}\label{wcomp}

Computed values 
of internal structure constants for main-sequence stars are 
available in the literature from other authors. We checked
the validity of our calculations by comparing our results of $\log k_2$ at the ZAMS
with those previously published ones. Our standard values of internal structure constants
are in qualitative agreement with those by H87, CG89a and 
CG92, as well as our distorted $\log k_2$ relative to those given
by C04, C05, C06 and C07.
%
\begin{figure}[htb]
\centering{
\includegraphics[width=8cm]{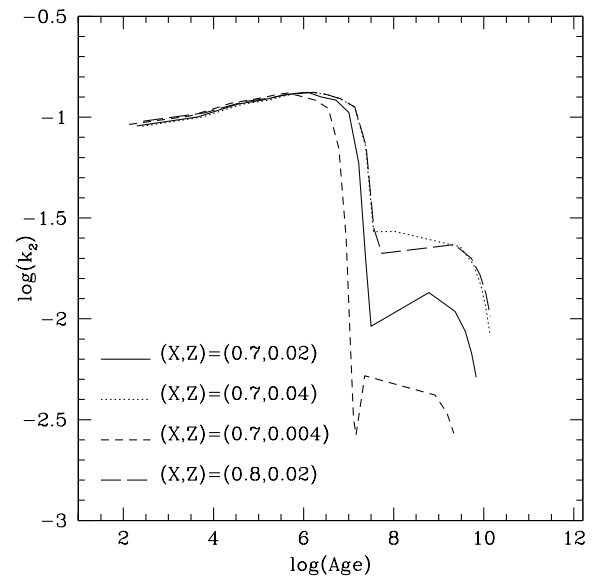}}
\caption{$\log k_2$ as a function of the stellar age
for a 1\,M$_{\odot}$ standard model, and for the initial chemical compositions
(X,\,Z)= (0.7,\,0.02), (0.07,\,0.04), (0.7,\,0.004), and (0.8,\,0.02).
}
\label{logk2qui}
\end{figure}
To better compare our standard 
results with the previous ones, we computed additional grids of 1\,M$_{\odot}$ standard models,
with different values of the mixing length parameter ($\alpha$=2.0 and $\alpha$=1.5)
and four different initial chemical compositions, (X,\,Z)=(0.7,\,0.02), (0.7,\,0.04),
(0.7,\,0.004), and (0.8,\,0.02). From Fig.\,\ref{logk2qui}, 
where the time-dependence of $\log k_2$ for $\alpha$=1.5 and
the initial chemical compositions above mentioned are shown,
one can have an insight on how the metallicity affects the value of
the second-order internal structure constant:
during the first 1\,Myr the chemical composition does not alter
$\log k_2$, but it becomes important from this age on; the metal-poorer
stars evolve to more centrally condensed configurations (lower values of
$\log k_2$) than their metal-richer counterparts. As it can also be seen from  
Fig.\,\ref{logk2qui}, using the initial chemical composition (X,\,Z)=(0.7,\,0.04)
(dotted lines) produces roughly the same effect on the evolution of $\log k_2$ as
using the initial chemistry (X,\,Z)=(0.8,\,0.02) (long-dashed lines).

In Table\,\ref{standcomp} we list the values at the ZAMS of the internal structure constants
obtained from our standard and distorted 1\,M$_{\odot}$ models and also from other works in the literature. 
The values for our standard models were obtained with the same initial chemical
compositions of Fig.\,\ref{logk2qui}; they can be 
compared with the values by, e.g.,
H87, CG89a and CG92. 

We start by noting that
for the initial chemical composition (X,\,Z)=(0.7,\,0.02) and $\alpha$=1.5, our 1\,M$_{\odot}$ model produces
a lower value of $\log k_2$ than that obtained by 
CG92, while for the same metallicity and $\alpha$=2.0, it 
produces a lower value of $\log k_2$ than those 
of CG89a and  
H87.
Now, let us consider the $\alpha$ that fits the sun for
each model, namely $\alpha$=1.5 for this work and CG92, and 
$\alpha$=2.0 for CG89a and H87. 
Still keeping the same metallicity, 
our models produce the most centrally
condensed 1\,M$_{\odot}$ star, followed by that of H87,  
CG89a and CG92, in this order.
{These differences between our values of $\log k_2$ and 
those by CG89a and CG92 can be attributed
to the use of different opacities, as they used Los Alamos Opacity Libraries and OPAL \citep{rogers3} opacities respectively,
while we used more recent OPAL opacities from
\citet{rogers1}; in addition, CG92 considered mass loss in their 
models.}
It is worth noting that although CG92 used more updated opacities
than CG89a, for the case of (X,\,Z)=(0.7,\,0.02) the former authors obtained
a value of $k_2$ for their 1\,M$_{\odot}$ model at the ZAMS
1.34 times greater than that obtained by the latter ones.
On the other hand, the CG92 models produce lower values of $k_2$
than those of CG89a for masses greater than 1.1\,M$_{\odot}$.

%
%
\begin{table}[htb]
\caption[The comparison between the available $\log$\,($k_2$).]
{Values of $\log k_2$ for a 1\,M$_{\odot}$ star model at the ZAMS,
obtained with our models (standard and distorted) and those previously published
in the literature.
} 
\vspace{0.4cm}
\label{standcomp}  
\vspace{-0.2cm}
\centering
\begin{tabular}{lccr}
\hline \hline
(X,\,Z)                         & $\alpha _{\rm MLT}$ & Reference              &  $\log (k_2)$ \\ \hline
\multirow{5}{*}{(0.7, 0.02)}    & 2.0                 & H87          &  $-$1.768     \\
                                & 2.0                 & CG89a        &  $-$1.747     \\
                                & 2.0                 & (standard)   &  $-$1.804     \\
                                & 1.5                 & CG92         &  $-$1.619     \\
                                & 1.5                 & (standard)   &  $-$1.975     \\ \hline
\multirow{3}{*}{(0.7, 0.04)}    & 2.0                 & H87          &  $-$1.614     \\
                                & 2.0                 & (standard)   &  $-$1.548     \\
                                & 1.5                 & (standard)   &  $-$1.628     \\ \hline
\multirow{3}{*}{(0.7, 0.004)}   & 2.0                 & H87          &  $-$2.035     \\
                                & 2.0                 & (standard)   &  $-$2.181     \\
                                & 1.5                 & (standard)   &  $-$2.281     \\ \hline
\multirow{3}{*}{(0.8, 0.02)}    & 2.0                 & H87          &  $-$1.578     \\
                                & 2.0                 & (standard)   &  $-$1.688     \\
                                & 1.5                 & (standard)   &  $-$1.776     \\ \hline
\multirow{2}{*}{(0.7, 0.02)}    & 1.68                & C04          &  $-$1.583     \\
                                & 1.68                & (standard)   &  $-$1.859     \\
                                & 1.68                & (distorted)  &  $-$1.895     \\ \hline
\multirow{2}{*}{(0.748, 0.004)} & 1.68                & C05          &  $-$1.840     \\
                                & 1.68                & (standard)   &  $-$2.112     \\
                                & 1.68                & (distorted)  &  $-$2.145     \\ \hline
\multirow{2}{*}{(0.73, 0.010)}  & 1.68                & C06b         &  $-$1.682     \\
                                & 1.68                & (standard)   &  $-$1.981     \\
                                & 1.68                & (distorted)  &  $-$2.004     \\ \hline
\multirow{2}{*}{(0.64, 0.04)}   & 1.68                & C07          &  $-$1.523     \\
                                & 1.68                & (standard)   &  $-$1.852     \\
                                & 1.68                & (distorted)  &  $-$1.866     \\ \hline
\end{tabular}
\end{table}
 
For the remaining initial chemical compositions,
we can only compare our 1\,M$_{\odot}$ standard model with those
by H87.
For the case of
$\alpha$=2.0, we notice that H87 obtained 
less mass-concentrated models
for (X,\,Z)=(0.7,\,0.004) and (X,\,Z)=(0.8,\,0.02) than us, while
for 
(X,\,Z)=(0.7,\,0.04) it is the opposite.
We can attribute
this fact to two reasons:
(1) although H87 obtained the lowest values of $k_2$ for one of the four metalicities
used in this comparison, he used older opacities from \citet{cox69}; 
and (2) our best fit to the sun is obtained
with $\alpha$=1.5, not with $\alpha$=2.0. 
Regarding this last point, if we turn our attention to the
models for which the value of $\alpha$ can reproduce the solar radius at the solar age 
($\alpha$=1.5 for this work and $\alpha$=2.0 for H87),
we see that our models again produce a more 
centrally condensed star than those of H87, independent of the chemistry.

Finally, we compare the $\log(k_2)$ values from our distorted models with
those of C04, C05, C06b and C07.
In order to perform suitable comparisons, we computed additional 1\,M$_{\odot}$ models with
the same physical inputs used in those works, namely $\alpha$=1.68 and initial
chemical compositions of (X,\,Z)=(0.7,\,0.02), (X,\,Z)=(0.748,\,0.004), (X,\,Z)=(0.730,\,0.010),
and (X,\,Z)=(0.64,\,0.04); the corresponding values of $\log k_2$
at the ZAMS are shown in the lower part of Table \,\ref{standcomp}.
We note that although the effects of tides and rotation are
also considered in the models by Claret and collaborators (\citealt{claret98};
Claret 2008, personal communication), their values presented in Table \,\ref{standcomp}
correspond to isolated, non-rotating stars; for comparison purposes, we also present in it our
values calculated for isolated, non-rotating stars with the same masses and chemical composition
as well as the corresponding ones including tidal and rotational distortions.
For all chemical compositions, our models again produce more mass-concentrated stars
than those by C04, C05, C06b, and C07.
For the mass range of this comparison, these differences in the $\log k_2$ values
can be due to the presence of distortion effects in our models and also to
the use of a different equation of state: while we use OPAL \citep{rogers2} and
\citet{mihalas} EOS, those authors use the CEFF \citep{dalsgaard} EOS.

\begin{figure}[htb]
\centering{
\includegraphics[width=8cm]{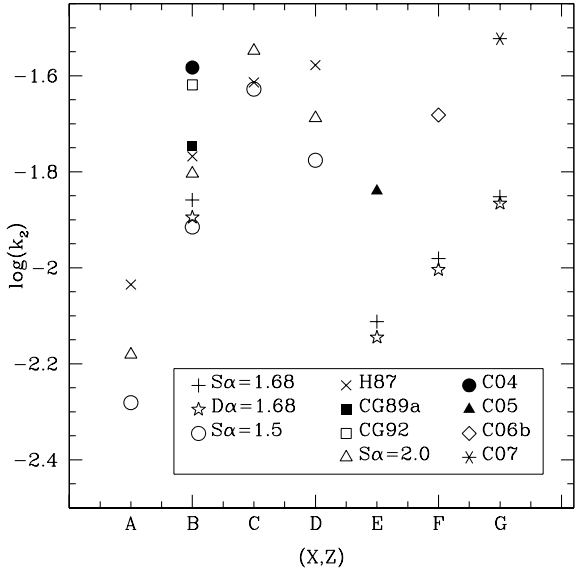}
}
\caption{A schematic comparison between $\log k_2$ values obtained with our 
standard models (S$\alpha$=1.5 and S$\alpha$=2.0), our distorted models
(D$\alpha$=1.68) and those obtained by other authors. 
The values plotted are the same as in Table\,\ref{standcomp}, and
the corresponding initial chemical compositions are the following, labeled
A to G respectively: (X,Z)=(0.7,\,0.004), (0.7,\,0.02), (0.7,\,0.04),
(0.8,\,0.02), (0.748,\,0.004), (0.73,\,0.01), and (0.64,\,0.04).
}
\label{comptabk2}
\end{figure}

Fig.\,\ref{comptabk2} presents a graphical comparison
between the values of $\log k_2$ obtained in this work and those obtained by 
H87, CG89a, CG92, C04, C05, C06b, and C07. This figure gives an easier overview of
Table\,\ref{standcomp}.

\section{Comparison between theory and observations}\label{comphteoobs}

Double-lined eclipsing binary systems are good candidates to test evolutionary
models, but very few systems are as adequate as EK~Cep (P=$4^{d}\!.42$).
The mass and radius of its primary component are 
M$_1$=$2.029\pm0.023$\,M$_{\odot}$ and
R$_1$=$1.579\pm0.007$\,R$_{\odot}$, while, for the secondary,
M$_2$=$1.124\pm0.012$\,M$_{\odot}$ and
R$_2$=$1.315\pm0.006$\,R$_{\odot}$ \citep{claret06a}. 
As mentioned by \citet{claret06a}, EK~Cep has accurate 
determination of absolute parameters, at least concerning 
masses and radii (see Table\,\ref{ekcpproper}); its secondary 
component is a pre-MS star,
the apsidal motion presented by the system has a high relativistic contribution, the less massive
component has its Lithium abundance determined observationally, 
the metallicity of the binary is evaluated, and, further, the 
rotational velocity of each component
is observationally measured.
Besides that, EK Cep is the only known pre-MS system with measured apsidal motion.
For all these reasons, we choose EK Cep to test our new models.
 
%
%
\begin{table}[htb]
\caption{Absolute dimensions of EK Cep \citep{claret06a,tomkin83}.} 
\label{ekcpproper}
\vspace{0.1cm}
\centering
\begin{tabular}{lrr}
\hline \hline
Parameter                   & Primary           & Secondary \\ \hline
Mass (M$_{\odot}$)          & 2.029 $\pm$ 0.023 &   1.124 $\pm$ 0.012 \\
Radius (R$_{\odot}$)        & 1.579 $\pm$ 0.007 &   1.315 $\pm$ 0.006 \\
$\log (g)$ (cgs)            & 4.349 $\pm$ 0.010 &   4.251 $\pm$ 0.006 \\
$\log$ (L/L$_{\odot}$)      & 1.17 $\pm$ 0.04   &   0.19 $\pm$ 0.07 \\
$\log$ (T$_{\rm eff}) (K)$  & 3.954 $\pm$ 0.010 &   3.756 $\pm$ 0.015 \\ 
v$_{\rm rot}$ (km/s) & 23$\pm$2 & 10.5$\pm$2 \\
\hline
\end{tabular}
\end{table}

EK Cep was discovered as an eclipsing binary by \citet{stroh59} from photographic
observations.
Photometric elements and
a spectroscopic orbit of the system were first presented by
\citet{ebbi66a,ebbi66b}, and further revised by other authors.
The apsidal motion of EK Cep was first reported in \citet{khaliu83}.
\citet{tomkin83} determined the masses and the radii
for its primary and secondary components. He also noted that the secondary is oversized in
comparison with a main-sequence star with the same mass and supposed that it might still be contracting
towards the main sequence. \citet{hill84}
concluded that both components are zero-age main-sequence objects, despite
\citet{tomkin83} suggestion about the secondary.
\citet{gimenez85} obtained a good agreement between theoretical
and observationally determined apsidal motion rates.
\citet{popper87} reported some anomalies in the secondary of EK~Cep, such
as low effective gravity and temperature as well as the excess radiation in the blue band,
that appear to be consistent with the hypothesis of the pre-MS nature of this star.
From high-resolution spectroscopy in the LiI\,$\lambda$6708\,\AA\ region,
\citet{martin93} determined the lithium abundance of EK~Cep B
and provided new evidence that it has not settled onto the ZAMS.
\citet{claret95b} compared the observed parameters of EK~Cep with theoretically
predicted values; they derived a common age for the system around 2$\times$10$^7$\,yr and
confirmed the fainter component as a pre-MS star, while the more massive companion is
in the beginning of the Hydrogen-burning phase.
They estimated a Newtonian apsidal motion rate that is in agreement with 
the observations, considering a predicted relativistic contribution of 
about 40\%. Those authors also found that the lithium
depletion computed by their models is consistent 
with the abundances determined by \citet{martin93}.

The evolutionary status of EK Cep was studied by other authors.
\citet{yildiz03} modeled the component stars by invoking
a rapidly rotating core for the primary. \citet{marques04} investigated the role
of overshooting on the modeling of pre-MS evolution of the secondary.
\citet{claret06a} presented the most recent analysis about EK~Cep;
due to problems with the empirical determination of the effective
temperatures of the component stars, he adopted the effective temperature 
ratio (TR=T$_{\rm eff,2}$/T$_{\rm eff,1}$), which
is better determined from the light curve analysis than their absolute values.
Inconsistencies found in the photometric distances for both components support
this approach.
\citet{claret06a} used other constraints in his
analysis as the radii, apsidal motion and lithium depletion.
With a rotating model
(assuming local conservation of the angular momentum), $\alpha$=1.4
and (X,\,Z)=(0.7075,\,0.0175), he fitted the radii and TR
in the same isochrone (24.2$\times 10^6$ yr).

\begin{figure}[t]
\centering{
\includegraphics[width=8cm]{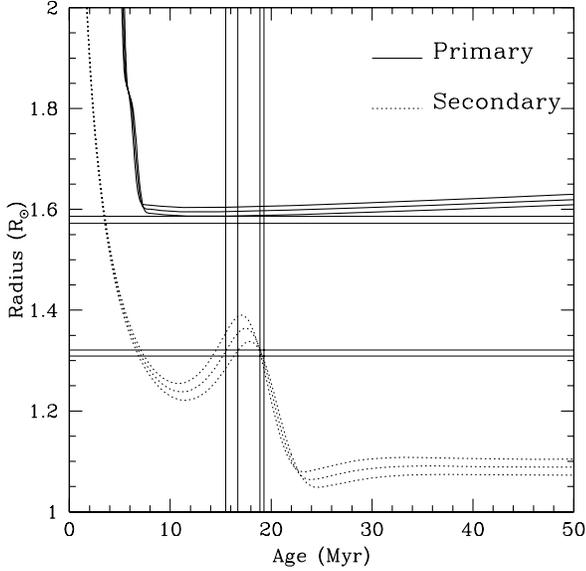}
}
\caption{Radii predicted by our binary rotating models
for EK Cep components determined masses. The additional tracks refer to
the maximum and minimum mass of each component, according to their
errors. 
Full lines denote the primary while dotted ones denote the
secondary. Note that there is an acceptable agreement between radii
and effective temperature ratio (see Fig.\,\ref{ektef}) for two
age intervals (vertical lines). Horizontal lines
represent the error bars in radii determinations.}
\label{ekrad}
\end{figure}

\begin{figure}[!t]
\centering{
\includegraphics[width=8cm]{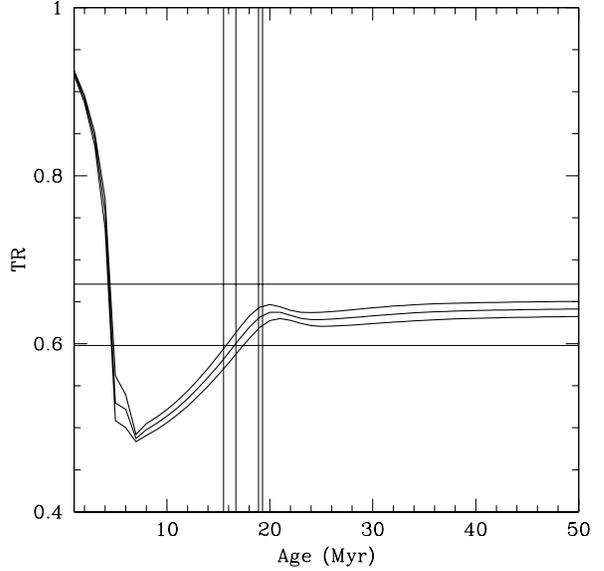}
}
\caption{Effective temperature ratio (TR=T$_{\rm eff,2}$/T$_{\rm eff,1}$)
predicted by our binary rotating models
for EK Cep components determined mass. The additional tracks refer to
the maximum and minimum mass of each component star, according to the
errors in their determination. The vertical lines correspond to the same age intervals as in Fig.\ref{ekrad}.
Horizontal lines represent the error bars in TR determinations.}
\label{ektef}
\end{figure}

As emphasized by \citet{claret06a}, before computing stellar models for a
given star it is fundamental to define clearly which observational parameters
will be used as constraints. Following the same approach as \citet{claret06a}, we adopt the masses,
radii and TR, leaving the apsidal motion, 
rotational velocities and lithium depletion to be used as additional
constraints after obtaining an acceptable solution. We then computed rotating binary
models with metallicities as close as possible to the solar metallicity,
following the conclusion by \citet{martin93} that the EK Cep secondary has
a metal contents typical of an young
disk, solar-type star.
Using an initial chemical composition of (X,\,Z)=(0.67,\,0.017)
and the mixing length parameter $\alpha$=1.5, we obtained
a model that reproduces the radii and TR for
EK Cep within the uncertainties, as shown in 
Figs.\,\ref{ekrad} and \ref{ektef},
deriving two possible age intervals for the system, 15.5-16.7$\times$10$^6$ and
18.9-19.3$\times$10$^6$~yr.
The input values we used are different from those by 
\citet{claret06a} since we are using different
stellar models, with different physics. By using his inputs, i.e. 
(X,\,Z)=(0.07075,\,0.0175) and $\alpha=1.4$, we were not able to reproduce, simultaneously,
both stellar radii at the same age interval; this is due to the fact that each model
needs slightly different inputs for adjusting to a same set of observed data.
%
%
%
\begin{figure}[tb]
\centering{\includegraphics[width=8cm]{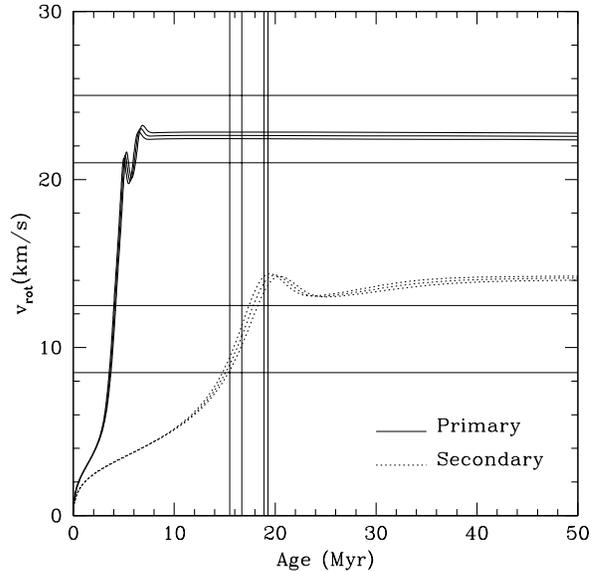}}
\caption{Time evolution of rotational velocities along the
pre-MS for EK Cep determined masses. 
Horizontal lines represent the error bars in rotational velocity 
determinations. Curve triplets and vertical lines defined as in Fig.\,\ref{ekrad}.}
\label{ekcpvel}
\end{figure}

According to our radii and TR analysis, 
the two derived age intervals allow
a reasonable adjustment of the observational data; we used 
additional data (rotational velocities of each component, lithium depletion
and apsidal motion constants) to constrain even more the system's age estimate.

We first tried to reproduce the observed rotational velocites of
EK Cep primary and secondary components at at least one of the derived age
intervals; in order to do so, we needed to use an initial angular momentum for
each component lower than that prescribed by Eq.\,\ref{kaweq}.
In Fig.\,\ref{ekcpvel} we show the time evolution of the rotational
velocities for EK Cep determined masses;
the vertical lines correspond to the two age intervals that
fit simultaneously the stellar radii and temperature ratio, while the
horizontal ones delimit the observed EK Cep rotational velocities. From that figure, one can see that
the older age interval does not 
match the observed rotational velocity of the secondary star.

%
%
\begin{figure}[tb]
\centering{
\includegraphics[width=8cm]{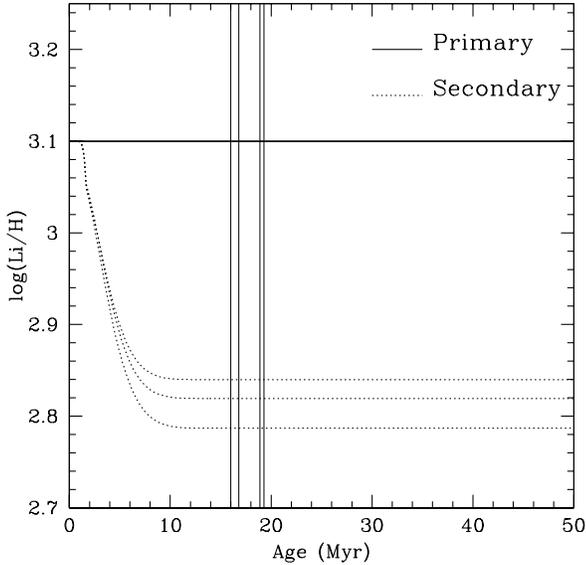}
}
\caption{Time evolution of lithium abundances for EK Cep determined masses
according to our rotating 
binary models. Curve triplets and vertical lines defined as in Fig.\,\ref{ekrad}.
Note that the curves corresponding to the primary mass and to its error bar collapse
to a single horizontal line.}
\label{ekli}
\end{figure}
%
%
\begin{figure}[htb]
\centering{
\includegraphics[width=8cm]{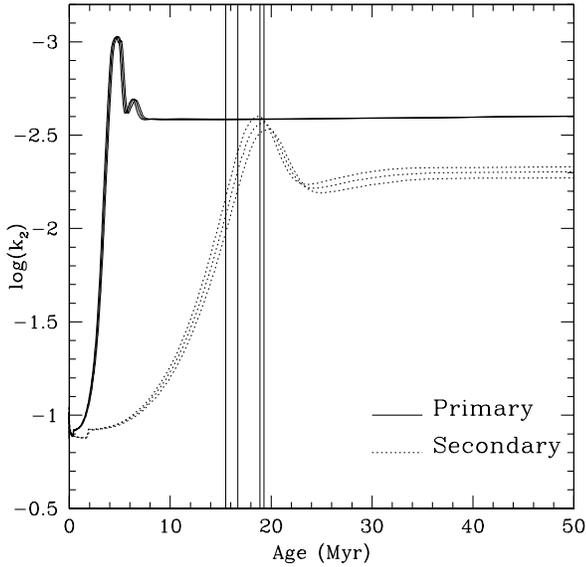}
}
\caption{The second-order apsidal motion constant as a function of the stellar age for EK Cep,
as obtained from our rotating binary models.
Curve triplets and vertical lines defined as in Fig.\,\ref{ekrad}}.
\label{ekk2}
\end{figure}
The lithium depletion of our rotating binary models is shown in
Fig.\,\ref{ekli}, in which the Li contents is plotted against
the stellar age. 
We started from an initial
lithium abundance of $\log$ (Li/H)=3.1 \citep[as][]{franca2}.
As expected, we do not find any depletion for the primary during its pre-MS evolution due to its higher mass; for the secondary,
we find a depletion of about 0.28 dex in both possible age intervals for EK Cep.
This corresponds to a lithium abundance of $\log$~(Li/H)=2.82$\pm$0.03,
consistent with the surface value of $\log$~(Li/H)=3.1$\pm$0.3
measured by \citet{martin93}.
The lithium depletion does not allow us
to choose between the two possible age intervals for the system, 
since it does not vary in the whole interval from  10\,Myr to  50\,Myr.

Finally, we investigated the apsidal motion rate of EK~Cep with
the same rotating binary models. The values for the orbital eccentricity
($0.109\pm 0.003$ degrees) and the orbital inclination ($89.3\pm 0.1$\,degrees)
were taken from \citet{petrova99}. The anomalistic period of EK Cep
is $P_{\rm an}\,=\,4.4278062\pm 0.0000005$\,days \citep{claret06a}.
In Fig.\,\ref{ekk2} we show the variation of $\log (k_2)$
for the observed masses (with their errors) as a function of age.
As before, the vertical lines indicate the age intervals
that fit both the radii and effective temperature ratio of EK Cep components, and from which we 
took the values of the internal structure constants to derive the apsidal motion rate.

For the first age interval,
15.5-16.7\,Myr (AI1),
our models result in $\rm{k_2}$ values of 
0.00261$\pm$0.00001 and 0.007$\pm$0.003 for the primary
and the secondary components, respectively.
From Eq.\,(\ref{k2theor}) we obtain $\log \bar{k}_{2\,\rm theo}$=$-$2.3$\pm$0.2 as the mean value of 
the second-order apsidal motion constant of the system.
The Newtonian contribution of the apsidal motion rate is obtained from Eqs.\,(\ref{dotomega}) and (\ref{k2theor}), resulting in 
$\dot \omega _N$=(0.00033$\pm$0.00011)\,$^{\circ}$/cycle.
We also calculated the relativistic contribution of the advance of periastron,
$\dot \omega _R$=(0.000434$\pm$0.000005)\,$^{\circ}$/cycle, which, based on our
models, corresponds to about 49-68\% of the total rate.
From these values, our models result in an apsidal motion rate of
$\dot \omega$=(0.00076$\pm$0.00012)\,$^{\circ}$/cycle and, consequently, an apsidal period of
U$_{\rm theo}$=(5800$\pm$800)\,yr; these values
are comparable with the observed ones.
However, for the second age interval, 
18.9-19.3\,Myr (AI2), we were not able to reproduce the observed 
apsidal motion of EK Cep by using our rotating binary models.
As can be seen in Table\,\ref{tabaps}, the apsidal motion quantities
we found in AI2 ($\log k_2$, apsidal period $U$ and apsidal motion rate 
$\dot\omega$) are not in agreement with the observed values, even
considering the error bars.
%
%
\begin{table}[tb]
\caption[Summary of the quantities related to the apsidal motion.]
{Summary of the quantities related to the apsidal motion.}
\label{tabaps}
\vspace{0.1cm}
\centering
\advance\tabcolsep by -3pt
\begin{tabular}{lrrr}
\hline \hline
                       & $\log k_2$       & U (years)    & $\dot\omega$ ($\circ$/cycle)  \\ \hline
Our models (AI1) & $-$2.3  & 5800 & 0.00076       \\[-4pt]
                            & $\pm$2  & $\pm$800    & $\pm$12 \\
Our models (AI2) & $-$2.57 & 7070 & 0.00061       \\ [-4pt]
                            & $\pm$2  & $\pm$50  & $\pm$1       \\
Observed                    & $-$2.09 & 4500     & 0.00097  \\ [-4pt]
                            & $\pm$9  & $\pm$700 & $\pm$15 \\
\citet{claret06a}           & $-$2.116& 4600     & 0.00095  \\ [-4pt]
                            & $\pm$6  & $\pm$400 & $\pm$8 \\ \hline
\end{tabular}
\end{table}

\citet{claret06a} used results of a series of works
(\citealt{khaliu83b,gimenez85b,hill84,claret95b}; and others)
to derive the observed apsidal motion rate of this system. He
reported the observed mean value of the internal structure constant as
$\log \bar{k}_2$=$-$2.09$\pm$0.09, which is equivalent to an observed
apsidal motion of about  $\dot \omega _{\rm obs}$=(0.00097$\pm$0.00015)\,$^{\circ}$/cycle, 
after applying the relativistic correction. This apsidal motion rate
produces an apsidal period of U$_{\rm obs}$=(4500$\pm$700)\,years.
From Table~\ref{tabaps}, where our results, those by \citet{claret06a}
and the observed determinations are summarized, it can
be seen that our predicted values for AI1
are in good agreement with the observed ones 
\citep[as well as those by][]{claret06a}, and the differences
lie within the errors.

At first glance, we could be tempted to conclude
that the second age interval AI2 gives a better agreement
for the EK Cep system than the first one due to the following reasons:
(1) at AI2, the radius of the secondary component constrains the age of
the system to a much narrower age interval, as shown in Fig.\,\ref{ekrad}; and
(2) the TR curves produced by our models lie very well inside the AI2 age interval,  
but are fitted just marginally for the earlier AI1 age interval (Fig.\,\ref{ektef}).
However, in that age interval our rotating binary models were not able to 
reproduce neither the rotational velocity of the secondary star 
(Fig.\,\ref{ekcpvel}) nor the apsidal motion rate for the system
(Table\,\ref{tabaps}), while the opposite did happen for AI1. Therefore,
AI1 seems to be more suitable than AI2 as the age interval 
for investigating the evolutionary status of EK Cep.
For these reasons, with regard to our models,
we suggest that the age of EK Cep is between 15.5 and 16.7\,Myr.

\section{Conclusions}\label{apsconc}

We computed stellar evolutionary models that take into account the
combined effects of rotation and of tidal forces due to a companion star,
obtaining values of the internal structure constants for
low-mass, pre-MS stars.
Our approximation for binary model calculations
(with effects of tidal forces) do not include the pseudo-potential
centrifugal terms arising from the orbital motion of the components
around the center of mass of the system.
For all sets of models, namely standard, binary, rotating and rotating binary 
models, we tabulated the internal structure constants and the gyration radii 
for ZAMS models and their time evolution. 
Distorted models result in more mass-concentrated stars and produce larger 
gyration radii at the ZAMS than standard ones.

The non-standard evolutionary tracks are cooler than their standard 
counterparts, mainly for low-mass stars. 
Regarding the internal structure of our stellar
models, we verified that tidal effects act in the same way as rotational ones
but in a smaller scale; the relative importance of these effects
on the apsidal motion constants depend mainly on the orbital separation and 
the star's rotation rate. Besides, rotationally and tidally distorted models
produce some effects in the physical quantities of a star, in comparison with the
standard models, that are in one direction, for masses below $\sim$0.7\,M$_{\odot}$,
and in the opposite direction, for masses above this threshold.
Though \citet{sack70} associated this behavior in 
distortion effects with the transition from the p-p chain to the CNO cycle,
this transition occurs in a mass range about 1.3-1.5\,M$_{\odot}$,
considerably above the threshold we observed with our models for the change of behavior
of the physical quantities of the stars.

We also found that, for masses lower than 0.5\,M$_{\odot}$, the relative 
importance of the second-order internal structure constants over those of higher
order is much lower than for the mass interval of 1.5-3.8\,M$_{\odot}$; 
hence, in the low-mass range, the usual assumption that the harmonics of order
greater than $j$=2 can be neglected seems not to be entirely justified
except maybe for computing the apsidal motion rate
$\dot{\omega}$.

Our results on internal structure constants
were compared with those available in the literature and found to be compatible 
with them. The $k_2$ values obtained from our standard
models are smaller than those last published by 
H87, CG89a and CG92, except for a given model by H87 with the inputs 
(X,\,Z)=(0.7,\,0.04) and $\alpha$=2.0.
Our rotating binary models produce internal structure
constants even smaller, resulting in more mass-concentrated configurations
than the models by C04, C05, C06, and C07.
These comparisons were made for
representative 1\,M$_{\odot}$ models, with the same initial chemical compositions of those works.

Using our set of evolutionary tracks for rotating binary
models, we also investigated the evolutionary status of the interesting
double-lined eclipsing binary system EK Cep. Its primary, a 2.029
M$_{\odot}$ star, seems to be in the hydrogen-burning phase, and its secondary,
a solar-like star (1.124 M$_{\odot}$), is confirmed as a pre-MS star.
By using a model with an initial chemical composition of
(X,\,Z)=(0.67,\,0.017) and a mixing length 
parameter of $\alpha$=1.5, we reproduced stellar radii and the effective 
temperature ratio of EK Cep in two different age intervals. 
We also followed the lithium contents during the pre-MS evolution of
both components; as expected, we do not find any significant depletion
for the primary, while the Li depletion for the secondary agrees with
the observed values within the uncertainties for both age intervals. 
However, for the later age interval of 18.9-19.3\,Myr, we were not able to fit neither
the observed rotational velocity of both components nor the observed
apsidal motion rate of the system. On the other hand, all observed
quantities used in our analysis were reproduced in the earlier age interval
of 15.5-16.7\,Myr. Although being a broader range for the age of EK~Cep
and introducing greater error bars to the apsidal motion
quantities, it seems to give a more confident age for the system. 

In this paper, we presented a first attempt to introduce the combined
structural effects of tides and rotation in the \texttt{ATON} evolutionary
code. Even with simple assumptions about the secondary component of a binary system, we could
verify the importance of those combined effects on stellar structure and evolution. Work is in
progress, where we remove the assumption that the disturbing star can be 
treated as a point mass, include the pseudo-potential
centrifugal terms caused by the orbital motion
and consider the simultaneous evolution of the 
two components of a binary system.

\begin{acknowledgements}
The authors thank Drs. Francesca D'Antona (INAF-OAR, Italy) and Italo Mazzitelli
(INAF-IASF, Italy) for granting them full access to the \texttt{ATON} evolutionary code.
We also are grateful to the referee, Dr.~B. Willems, for his many
comments and suggestions that helped to improve this work.
The financial
support from the Brazilian agencies CAPES, CNPq and FAPEMIG is also acknowledged.
\end{acknowledgements}


\begin{thebibliography}{} 

   \bibitem[Alexander \& Ferguson, 1994]{alexander} Alexander, D.R., 
           Ferguson, J.W.\ 1994, ApJ, 437, 879
   
   \bibitem[Batten, 1973]{batten73} Batten, A.H.\ 1973, in {\it Binary and 
           Multiple Systems of Stars}, Pergamon Press, Oxford

   \bibitem[B\"ohm-Vitense, 1958]{bohm} B\"ohm-Vitense, E.\ 1958, 
           Z. Astroph., 46, 108

   \bibitem[Brooker \& Olle, 1955]{brooker55} Brooker, R.A., Olle, T.W.\ 
           1955, \mnras, 115, 101

   \bibitem[Canuto et al., 1996]{canuto96} Canuto V.M., Goldman I., Mazzitelli I.,
           1996, ApJ, 473, 550
   \bibitem[Carson, 1976]{carson76} Carson, T.R.\ 1976, \araa, 14, 95

   \bibitem[Caughlan \& Fowler, 1988]{caughlan} Caughlan, G.R., 
           Fowler, W.A.\ 1988, Atomic Data Nucl. Tab., 40, 283

   \bibitem[Chaboyer et al., 1995]{chaboyer95} Chaboyer, B., Demarque, P, 
            Pinsonneault, M.H.\ 1995, ApJ, 441, 865
    
   \bibitem[Chandrasekhar, 1933]{chandra33} Chandrasekhar, S.\ 
           1933, \mnras, 93, 390

   \bibitem[Christensen-Dalsgaard \& D\"appen, 1992]{dalsgaard} Christensen-Dalsgaard, J.,
           D\"appen, W., 1992, \aapr 4, 267

   \bibitem[Claret, 1995]{claret95a} Claret, A.\ 1995a, A\&AS, 109, 441

   \bibitem[Claret, 1998]{claret98}Claret, A.\ 1998, A\&AS, 131, 195
   
   \bibitem[Claret, 1999]{claret99} Claret, A.\ 1999, A\&A, 350, 56

   \bibitem[Claret, 2004]{claret04}Claret, A.\ 2004, A\&A, 424, 919 (C04)

   \bibitem[Claret, 2005]{claret05}Claret, A.\ 2005, A\&A, 440, 647 (C05)
   
   \bibitem[Claret, 2006a]{claret06a} Claret, A.\ 2006a, A\&A, 445, 1061

   \bibitem[Claret, 2006b]{claret06b} Claret, A.\ 2006b, A\&A, 453, 769 (C06b)

   \bibitem[Claret, 2007]{claret07} Claret, A.\ 2007, A\&A, 467, 1389 (C07)

   \bibitem[Claret \& Gim\'enez, 1989a]{claret89a} Claret, A., Gim\'enez, A.\ 
           1989a, \aaps, 81, 1 (CG89a)

   \bibitem[Claret \& Gim\'enez, 1989b]{claret89b} Claret, A., Gim\'enez, A.\ 
           1989b, \aaps, 81, 37

   \bibitem[Claret \& Gim\'enez, 1991]{claret91} Claret, A., Gim\'enez, A.\ 
           1991, \aaps, 87, 507

   \bibitem[Claret \& Gim\'enez, 1992]{claret92} Claret, A., Gim\'enez, A.\ 
           1992, \aaps, 96, 255 (CG92)
   
   \bibitem[Claret \& Gim\'enez, 1993]{claret93} Claret, A., Gim\'enez, A.\ 
           1993, \aap, 277, 487

   \bibitem[Claret \& Gim\'enez, 2001]{claret01} Claret, A., Gim\'enez, A.\ 
           2001, LNP, 563, 1

   \bibitem[Claret et al., 1995b]{claret95b} Claret, A., 
           Gim\'enez, A., Mart\'{\i}n, E.L.\ 1995b, A\&A, 302, 741

   \bibitem[Claret \& Willems, 2002]{claret02} Claret, A., Willems, B.\ 
           2002, \aap, 388, 518

   \bibitem[Cox \& Stewart, 1969]{cox69} Cox,\,A.N.,\,Stewart,\,J.N.\,1969,
           Nautshnij\,Informatsij, 15,\,1

   \bibitem[Cowling, 1938]{cowling38} Cowling, T.G.\ 1938, \mnras, 98, 734
   
   \bibitem[D'Antona \& Montalb\'an, 2003]{franca2} D'Antona, F., 
           Montalb\'an, J.\ 2003, A\&A, 412, 213

   \bibitem[Ebbighausen, 1966a]{ebbi66a} Ebbighausen, E.G.\ 1966a, AJ, 71, 642

   \bibitem[Ebbighausen, 1966b]{ebbi66b} Ebbighausen, E.G.\ 1966b, AJ, 71, 730

   \bibitem[Endal \& Sofia, 1976]{endal76} Endal, A.S., Sofia, S.\ 1976, ApJ, 210, 184

   \bibitem[Fliegener \& Langer, 1995]{fliegner95} Fliegner, J., 
            Langer, N.\ 1995, in: Wolf-RAyet Stars: Binaries, Colliding Winds,
            Evolution, IAU Symp. 163, K.A. van der Hucht and 
            P.M. Willians (eds), Kluwer
  
   \bibitem[Gim\'enez, 1985]{gimenez85} Gim\'enez, A.\ 1985, \apj, 297, 405

   \bibitem[Gim\'enez \& Margrave, 1985]{gimenez85b} Gim\'enez, A., 
           Margrave, T.E.\ 1985, AJ, 90, 358

   \bibitem[Hadjidemetriou, 1967]{hadjid67} Hadjidemetriou, J.\ 1967, 
           Adv.\ Astr.\ Astrophys., 5, 131


   \bibitem[Hejlesen, 1987]{hejlesen87} Hejlesen, P.M.\ 1987, \aaps, 69, 251 (H87)

   \bibitem[Heger et al., 2000]{heger00} Heger, A., Langer, N., Woosley, S.E., 2000,
            \apj, 528, 368 

   \bibitem[Hilditch, 2001]{hilditch01} Hilditch, R.W., 2001, 
           in {\it An Introduction to Close Binary Stars}, Cambridge
           University Press

   \bibitem[Hill \& Ebbighausen, 1984]{hill84} Hill, G., 
           Ebbighausen, E.G.\ 1984, AJ, 89, 1256

   \bibitem[Iglesias \& Rogers, 1993]{rogers1} Iglesias, C.A., 
           Rogers, F.J.\ 1993, ApJ, 412, 752

   \bibitem[Jeffery, 1984]{jeffery84} Jeffery, C.M.\ 1984, \mnras, 207, 323

   \bibitem[Khaliullin, 1983a]{khaliu83} Khaliullin, Kh.F.\ 1983a, 
           AZh., 60, 72

   \bibitem[Khaliullin, 1983b]{khaliu83b} Khaliullin, Kh.F.\ 1983b, 
           Sov.\ Astron., 27, 43

   \bibitem[Kawaler, 1987]{kawaler87} Kawaler, S.D.\ 1987, PASP, 99, 1322

   \bibitem[Keller \& Meyerott, 1955]{keller55} Keller, G., 
           Meyerott, R.E.\ 1955, \apj, 122, 32

   \bibitem[Kippenhahn \& Weigert, 1994]{kip:94} Kippenhahn, R., 
           Weigert, A. 1994, {\it Stellar Structure and Evolution},
           Springer-Verlag

   \bibitem[Kippenhahn \& Thomas, 1970]{kippen70} Kippenhahn, R., 
           Thomas, H.-C.\ 1970, in {\it Stellar Rotation}, 
           ed. A. Slettebak (Dordrecht: Reidel) (KT70)

   \bibitem[Kopal, 1959]{kopal59} Kopal, Z.\ 1959, in 
           {\it Close Binary Systems}, (New York: Wiley)

   \bibitem[Kopal, 1960]{kopal60} Kopal, Z.\ 1960, in
           {\it Figures of Equilibrium of Celestial Bodies},
           The University of Wisconsin Press

   \bibitem[Kopal, 1972]{kopal72} Kopal, Z.\ 1972, 
           Adv.\ Astron.\ Astrophys., 9, 1

   \bibitem[Kopal, 1974]{kopal74} Kopal, Z.\ 1974, 
           Astrophys.\ Space Sci., 27, 389

   \bibitem[Kopal, 1978]{kopal78} Kopal, Z.\ 1978, in {\it Dynamics
           of Close Binary Systems}, Reidel, Dordrecht

   \bibitem[Kopal, 1989]{kopal89} Kopal, Z.\ 1989, in {\it The Roche Problem}, 
           Kluwer Academic Publishers, Dordrecht

   \bibitem[Kushawa, 1957]{kushawa57} Kushawa, R.S.\ 1957, \apj, 125, 242

   \bibitem[Landin et al., 2006]{landin06} Landin, N.R., Ventura, P.,
           D'Antona, F., Mendes, L.T.S., \& Vaz, L.P.R.\ 2006, A\&A, 456, 269

   \bibitem[Law, 1980]{law80} Law W.-Y.,\ 1980, Ph.D. Thesis, Yale University

   \bibitem[Levi-Civita, 1937]{levi37} Levi-Civita, T.\ 1937, 
           Amer. J Math., 59, 225

   \bibitem[Marques et al., 2004]{marques04} Marques, J.P., Fernandes, J., 
           Monteiro, M.J.P.F.G.\ 2004, A\&A, 422, 239

   \bibitem[Mart\'{\i}n \& Claret, 1996] {claret96} Mart\'{\i}n, E.L., 
           Claret, A.\ 1996, \aap, 306, 408

   \bibitem[Mart\'{\i}n \& Rebolo, 1993]{martin93} Mart\'{\i}n, E.L., 
           Rebolo, R.\ 1993, A\&A, 274, 274

   \bibitem[Martynov, 1948]{martynov48} Martynov, D.Ya.\ 1948, 
           Izv.\ Engelhardt Obs., No. 25

   \bibitem[Martynov, 1973]{martynov73} Martynov, D.Ya.\ 1973, in Eclipsing 
           Variable Stars, V.P.\ Tsesevich (ed.), IPST 
           Astrophys.\ Library, Jerusalem

   \bibitem[Mendes et al., 1999a]{mendes} Mendes, L.T.S., D'Antona, F., 
           Mazzitelli, I.\ 1999a, A\&A, 341, 174

   \bibitem[Mendes, 1999b] {mendesphd} Mendes, L.T.S.\ 1999b, Ph.D. Thesis, 
           Federal University of Minas Gerais
  
   \bibitem[Meynet \& Maeder, 1997]{meynet97} Meynet, G., Maeder, A.\ 1997,
            A\&A, 321, 465

   \bibitem[Meynet \& Maeder, 2000]{meynet00} Meynet, G., Maeder, A.\ 2000,
            A\&A, 361, 101

   \bibitem[Maeder \& Meynet, 2000]{maeder00} Maeder, A., Meynet, G.\ 2000,
            A\&A, 361, 159

   \bibitem[Maeder \& Meynet, 2001]{maeder01} Maeder, A., Meynet, G.\ 2001,
            A\&A, 373, 555

   \bibitem[Meynet \& Maeder, 2002]{meynet02} Meynet, G., Maeder, A.\ 2002,
            A\&A, 390, 561

   \bibitem[Maeder \& Zahn, 1998]{maeder98} Maeder, A., Zahn, J.-P., 1998,
            A\&A, 334, 1000

   \bibitem[Mihalas et al., 1988]{mihalas} Mihalas, D., Dappen, W., 
           Hummer, D.G.\ 1988, ApJ, 331, 815 (M88)

   \bibitem [Mohan et al., 1990]{mohan90} Mohan, C., Saxena, R. M., 
            Agarwal, S. R.\ 1990, \apss, 163, 23

   \bibitem[Motz, 1952]{motz52} Motz, L.\ 1952, \aj, 115, 562

   \bibitem[Petrova \& Orlov, 1999]{petrova99} Petrova, A.V., 
           Orlov, V.V.\ 1999, AJ, 117, 587

   \bibitem[Pinsonneault, 1988]{pinson88} Pinsonneault, M.H.\ 1988, 
           Ph.D. Thesis, Yale University

   \bibitem[Pinsonneault et al., 1990]{pinson90} Pinsonneault, M.H.,
           Kawaler, S.D., Demarque, P.\ 1990, ApJS, 74, 501

   \bibitem[Popper, 1987]{popper87} Popper, D.M.\ 1987, ApJ, 313, 81

   \bibitem[Press et al., 1992] {nrecipes} Press, W.H., Teukolsky, S.A., 
           Vetterling, W.T., Flannery, B.P.\ 1992, Numerical Recipes in 
           Fortran 77: The art of Scientific Computing, Cambridge 
           University Press

    \bibitem [Rogers \& Iglesias, 1992] {rogers3} Rogers, F.J., 
            Iglesias, C.A.\ 1992, \apjs, 79, 507
 
   \bibitem[Rogers et al., 1996]{rogers2} Rogers, F.J., Swenson, F.J., 
           Iglesias, C.A.\ 1996, ApJ, 456, 902

   \bibitem[Ruci\'nski, 1969]{rucinski69} Ruci\'nski, S.M.\ 1969, 
           Acta Astron., 19, 125

   \bibitem[Ruci\'nski, 1988]{rucinski88} Ruci\'nski, S.M.\ 1988, 
           \aj, 95, 1895

   \bibitem[Russell, 1928]{russell28} Russell, H.N.\ 1928, \mnras, 88, 642

   \bibitem[Sackmann, 1970]{sack70} Sackmann, I.J.\ 1970, A\&A, 8, 76

   \bibitem[Sackmann \& Anand, 1969]{sack} Sackmann, I.J., 
           Anand, S.P.S.\ 1969, ApJ, 155, 257

   \bibitem[Sahade \& Wood, 1978]{sahade78} Sahade, J., Wood, F.B.\ 
           1978, in {\it Interacting Binary Stars}, Pergamom Press, Oxford

   \bibitem[Savonije \& Papaloizou, 1983]{savonije83} Savonije, G.J.,
           Papaloizou, J.C.B.\ 1983, MNRAS, 203, 581
 
   \bibitem[Savonije \& Witte, 2002a]{witte02a} Savonije, G.J.,
           Witte, M.G.\ 2002a, A\&A, 386, 211

   \bibitem[Sterne, 1939]{sterne39} Sterne, T.E.\ 1939, \mnras, 99, 662

   \bibitem[Strohmeier, 1959]{stroh59} Strohmeier, W.\ 1959, 
           {\it kleine Ver\"off. Bamberg\/}, No. 27

   \bibitem[Schwarzschild, 1958]{schwarz58} Schwarzschild, M.\ 1958, 
           in {\it Structure and Evolution of the Stars\/}, 
           Princeton Univ.\ Press

   \bibitem[Tomkin, 1983]{tomkin83} Tomkin, J.\ 1983, ApJ, 271, 717

   \bibitem[Ureche, 1976]{ureche76} Ureche, V.\ 1976, IAU Symp.\ No.\ 73, 
           P. Eggleton ed., Reidel, Dordrecht

   \bibitem[Ventura et al., 1998]{ventura98} Ventura, P., Zeppieri, A., 
           Mazzitelli, I., D'Antona, F.\ 1998, A\&A, 334, 953

   \bibitem[Yildiz, 2003]{yildiz03} Yildiz, M.\ 2003, A\&A, 409, 689

   \bibitem[Willems et al., 2003]{willems03I} Willems, B., van Hoolst, T., Smeyers, P.,
           2003, A\&A, 397, 973

   \bibitem[Willems \& Claret, 2003]{willems03} Willems, B., Claret, A., 
           2003, \aap, 410, 289

   \bibitem[Witte \& Savonije, 1999a]{witte99a} Witte, M.G.,
           Savonije, G.J.\ 1999a, A\&A, 341, 842

   \bibitem[Witte \& Savonije, 1999b]{witte99b} Witte, M.G.,
           Savonije, G.J.\ 1999b, A\&A, 350, 129

   \bibitem[Witte \& Savonije, 2001]{witte01} Witte, M.G.,
           Savonije, G.J.\ 2001, A\&A, 366, 840

   \bibitem[Witte \& Savonije, 2002b]{witte02b} Witte, M.G., 
           Savonije, G.J.\ 2002b, A\&A, 386, 222

   \bibitem[Zahn, 1966]{zahn66} Zahn, J.-P.\ 1966, AnAp., 29, 489

   \bibitem[Zahn, 1975]{zahn75} Zahn, J.-P.\ 1975, \aap,  41, 329
 
   \bibitem[Zahn, 1977]{zahn77} Zahn, J.-P.\ 1977, \aap,  57, 383

   \bibitem[Zahn, 1989]{zahn89} Zahn, J.-P.\ 1989, \aap,  220, 112

\end{thebibliography}
\end{document}